\documentclass[
superscriptaddress,
amsmath,
amssymb,
aps,
prx,
longbibliography,
twocolumn,
nonatbib 
]{revtex4-2}

\usepackage{tabularx}
\usepackage{graphicx}
\usepackage{dcolumn}
\usepackage{amsthm}
\usepackage{bm}
\usepackage{braket}
\usepackage{svg}
\usepackage{tikz}
\usepackage{ulem}
\usepackage{times}
\usepackage{xspace}
\usepackage{lineno}
\usepackage{extarrows}

\usetikzlibrary{quantikz}
\usepackage[colorlinks,linkcolor=blue,anchorcolor=blue,citecolor=blue]{hyperref}
\usepackage{xcolor}

\newtheorem{definition}{Definition}

\hypersetup{
	plainpages=true,
	breaklinks=true,%
	hypertexnames=false,
	pageanchor=true,
	colorlinks=true,
	linkcolor={red},
	citecolor={blue},
	urlcolor={blue},
	anchorcolor={black}
}

\usepackage{algorithm}
\usepackage{algorithmic}

\newcommand{\casql}{Laboratory of Quantum Information, School of Physics, University of Science and Technology of China, Hefei, Anhui, 230026, China}

\newcommand{\ceqi}{CAS Center For Excellence in Quantum Information and Quantum Physics, University of Science and Technology of China, Hefei, Anhui, 230026, China}

\newcommand{\aihf}{Institute of Artificial Intelligence, Hefei Comprehensive National Science Center, Hefei, Anhui, 230088, China}

\newcommand{\casss}{Key Laboratory of System Software (Chinese Academy of Sciences) and State Key Laboratory of Computer Science, Institute of Software, Chinese Academy of Sciences, Beijing 100190, China}

\newcommand{\origin}{Origin Quantum Computing, Hefei, Anhui, 230026, China}

\newcommand{\Agder}{Department of ICT and Center for AI Research, University of Agder (UiA), Jon Lilletuns vei 9, 4879 Grimstad, Norway}

\newcommand{\ahu}{Anhui University, Hefei, Anhui, 230039, China}

\begin{document}


\title{Experimental robustness benchmarking of quantum neural networks on a superconducting quantum processor}
	
	\author{Hai-Feng Zhang}
	\affiliation{\casql}
        \affiliation{\ceqi}
	
	\author{Zhao-Yun Chen}
    \email{chenzhaoyun@iai.ustc.edu.cn}
	\affiliation{\aihf}
	
	\author{Peng Wang}
	\affiliation{\casql}
	\affiliation{\ceqi}
 
	\author{Liang-Liang Guo}
	\affiliation{\origin}
	
	\author{Tian-Le Wang}
	\affiliation{\casql}
	\affiliation{\ceqi}
 
	\author{Xiao-Yan Yang}
	\affiliation{\casql}
	\affiliation{\ceqi}
 
	\author{Ren-Ze~Zhao}
	\affiliation{\casql}
	\affiliation{\ceqi}
 
	\author{Ze-An Zhao}
	\affiliation{\casql}
	\affiliation{\ceqi}
 
	\author{Sheng Zhang}
	\affiliation{\casql}
	\affiliation{\ceqi}
 
	\author{Lei Du}
	\affiliation{\origin}
 
	\author{Hao-Ran Tao}
	\affiliation{\origin}
	
	\author{Zhi-Long Jia}
	\affiliation{\origin}
	
	\author{Wei-Cheng Kong}
	\affiliation{\origin}
	
	\author{Huan-Yu~Liu}
	\affiliation{\casql}
	\affiliation{\ceqi}
 
	\author{Athanasios V. Vasilakos}
	\affiliation{\Agder}
		
	\author{Yang Yang}
	\affiliation{\aihf}
        \affiliation{\ahu}
 
	\author{Yu-Chun Wu}
	\affiliation{\casql}
        \affiliation{\ceqi}
	\affiliation{\aihf}
	
	\author{Ji Guan}
\email{guanji1992@gmail.com}
	\affiliation{\casss}
	
	\author{Peng Duan}
\email{pengduan@ustc.edu.cn}
	\affiliation{\casql}
	\affiliation{\ceqi}
 
	\author{Guo-Ping Guo}%
\email{gpguo@ustc.edu.cn}
	\affiliation{\casql}
	\affiliation{\ceqi}
	\affiliation{\origin}
    
	\begin{abstract}
        
Quantum machine learning (QML) models, like their classical counterparts, are intrinsically vulnerable to adversarial attacks, hindering their secure deployment. Here, we report the first systematic experimental benchmark of robustness for $20$-qubit quantum neural network (QNN) classifiers executed on a superconducting processor. Our benchmarking protocol features an efficient adversarial attack algorithm tailored for quantum hardware, enabling the diagnosis of QNN's robustness across diverse datasets. The empirical upper bound extracted from our attack experiments deviates by only $3 \times 10^{-3}$ from the analytical lower bound, providing strong experimental confirmation of our attack's precision and the tightness of the fidelity-based robustness bounds. Furthermore, our quantitative analysis reveals that adversarial training mitigates sensitivity to targeted perturbations by regularizing input gradients, thereby significantly enhancing QNN robustness. Additionally, we observe that experimentally measured QNNs exhibit higher adversarial robustness than classical neural networks, an effect attributed to inherent quantum noise. Our work establishes the first scalable and experimentally accessible framework for robustness benchmarking, paving the way for secure and reliable QML applications.

	\end{abstract}

	\maketitle

Over the past decade, machine learning (ML), particularly deep neural networks, has become indispensable to critical applications ranging from natural language processing to autonomous driving~\cite{vaswani2017attention,chib2023recent}. However, the widespread deployment of these models has concurrently brought their security vulnerabilities to the forefront. A pressing concern is their susceptibility to adversarial attacks--carefully crafted, often imperceptible input perturbations that can induce misclassification or erroneous behavior, posing risks in sensitive applications like facial recognition~\cite{madry2017towards,9252132,cao2024rethinking}. Rigorous robustness evaluation is thus imperative for ML reliability and safety~\cite{shafique2020robust}.

Quantum machine learning (QML)~\cite{Biamonte2017,cerezo2022challenges} seeks to leverage quantum computational principles for potential advantages in areas such as high-dimensional data analysis and complex system simulation simulation~\cite{lloyd2014quantum,Rebentrost2014,dunjko2016quantum,huang2022quantum}. Driven by advancements in quantum hardware platforms like superconducting circuits, the practical implementation and experimental investigation of quantum neural networks (QNNs) are gaining significant momentum~\cite{s41586-019-0980-2,Herrmann2022,Gong2023,Tacchino2019,Huang2021,huang2021quantum,zhang2024quantum,chen2025quantum}. Despite their distinct computational paradigm, QNNs unfortunately appear to inherit vulnerabilities analogous to their classical counterparts~\cite{Lu2020,Liu2020}. Therefore, ensuring the security and trustworthy deployment of QNNs, particularly their robustness against adversarial threats, has emerged as a critical research priority~\cite{franco2024predominant}. Initial theoretical work has formalized fidelity-based robustness bounds, enabling rapid estimation of robustness for specific input samples~\cite{Guan2020,lin2024veriqr}. In parallel, other researchers have explored the resilience of QMLs against black-box attacks transferred from classical domains, suggesting a potential quantum robustness advantage~\cite{West2023}.
	
\begin{figure*}[!t]
\centering
\includegraphics[width=0.95\textwidth]{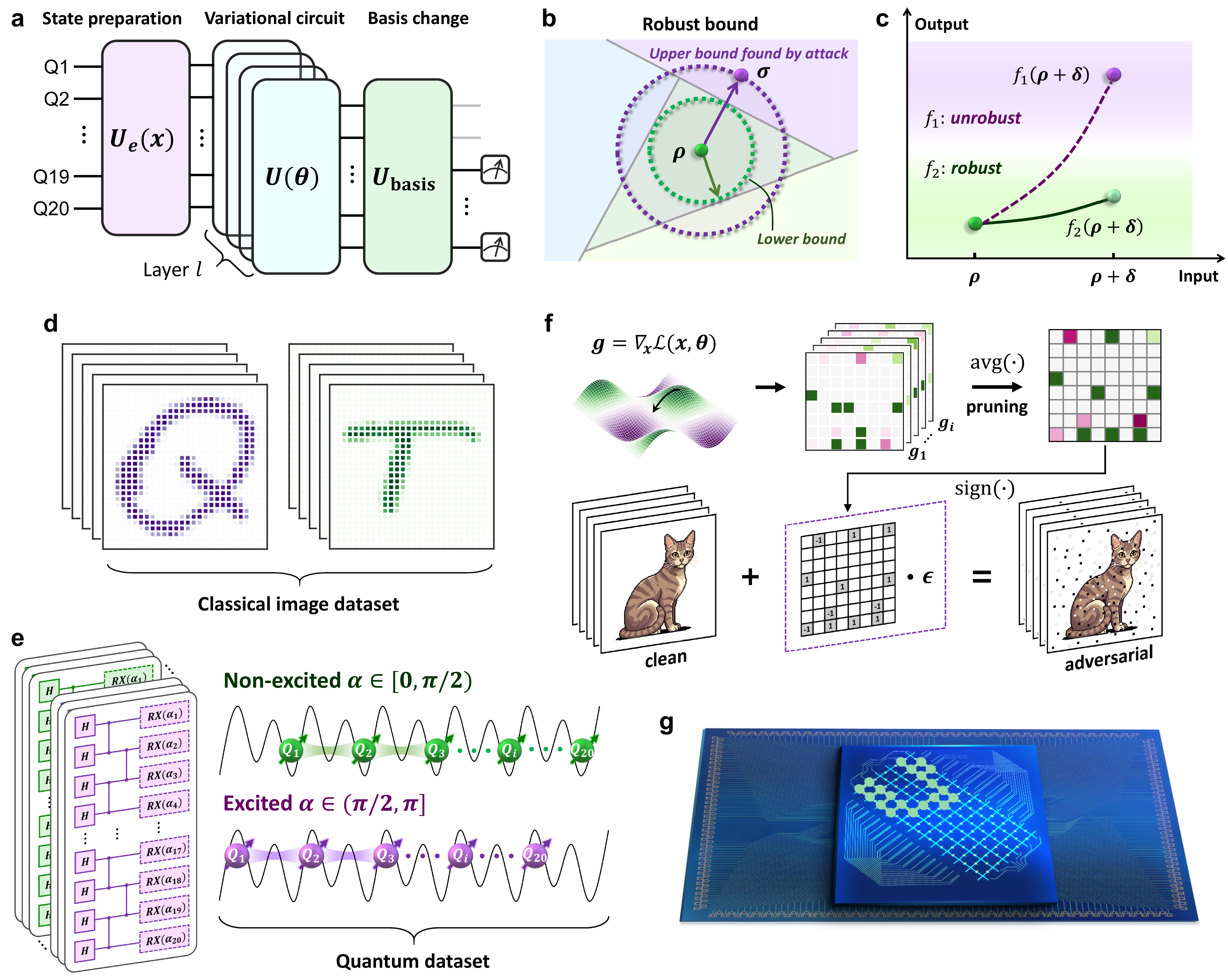}
\caption{\textbf{Experimental schematic for QNN robustness evaluation.} $\textbf{a}$, Architecture of the QNN classifier, consisting of a state preparation circuit, an $l$-layer variational circuit, and pre-measurement basis transformation gates. $\textbf{b}$, Illustration of the robustness lower and upper bounds of the classifier. $\textbf{c}$, Adversarial robustness, quantified by the output sensitivity to input perturbations. $\textbf{d}$, Visualization of handwritten letters ``Q'' and ``T'' used in the image classification. $\textbf{e}$, Quantum synthetic dataset for LCEI, illustrating the application of an $R_x(\alpha)$ after the linear cluster state, with states labeled as ``excited'' or ``non-excited'' based on the rotation angle $\alpha$. $\textbf{f}$, The proposed masked adversarial attack is characterized by identifying the vulnerable subspace through input-gradient analysis and applying localized perturbations. $\textbf{g}$, Schematic of the superconducting quantum processor, showing $72$ qubits and $126$ couplers in a 2D lattice, and $20$ qubits selected for the experiment are highlighted in green.}
\label{fig1}
\end{figure*}

To date, a systematic robustness evaluation of QMLs in experiments remains absent from the current literature. Achieving this requires not only a comprehensive set of metrics, but also hardware-efficient adversarial attack algorithms that can be implemented \textit{in-situ} on quantum processors. Rigorous robustness evaluation relies on adversarial examples generated from hardware, offering the sole faithful route to probing classifier susceptibility and benchmarking robustness bounds. Yet this endeavor is experimentally challenging at scale, as inherent quantum noise can obscure attack efficacy and distort robustness analyses. Consequently, quantitative validation of robustness improvements from methods like quantum adversarial training on physical systems has also remained elusive, with estimation predominantly qualitative~\cite{Ren2022}.

In this work, we overcome these challenges and propose an efficient adversarial attack algorithm tailored for noisy intermediate-scale quantum (NISQ) devices, enabling experimental benchmarking of robustness bounds for QNNs. In attack experiments on QNNs with up to 20 superconducting qubits, the extracted robustness upper bounds deviate from theoretical lower bounds by only a minimal margin (approximately $3 \times 10^{-3}$). This small discrepancy provides strong experimental validation that our attack strategy effectively identifies the vulnerable subspace of the QNN, while also confirming the tightness of fidelity-based robustness bounds. 

Additionally, building upon this attack strategy, we introduce a universal metric for quantifying adversarial robustness in quantum classifiers. In the largest-scale adversarial learning experiment, we successfully measured the robustness scores of QNNs across diverse datasets and training strategies. We also demonstrate that adversarial training substantially mitigates QNN sensitivity to adversarial perturbations through intrinsically regularizing input gradients, thereby enhancing robustness against target perturbations. Comparative experiments revealed an intriguing finding: Compared to classical neural networks, QNNs exhibit significantly higher adversarial robustness under identical tasks, which we attribute to robustness properties arising from gradient attenuation induced by inherent quantum noise. This work develops a scalable and experimentally realizable framework for robustness benchmarking, offering the first broadly applicable methodology for diagnosing the security and reliability of QMLs in realistic hardware settings.

\vspace{.5cm}
\noindent\textbf{\large{}Experimental setup and framework}{\large\par}

\begin{figure*}[!t]
\centering
\includegraphics[width=0.95\textwidth]{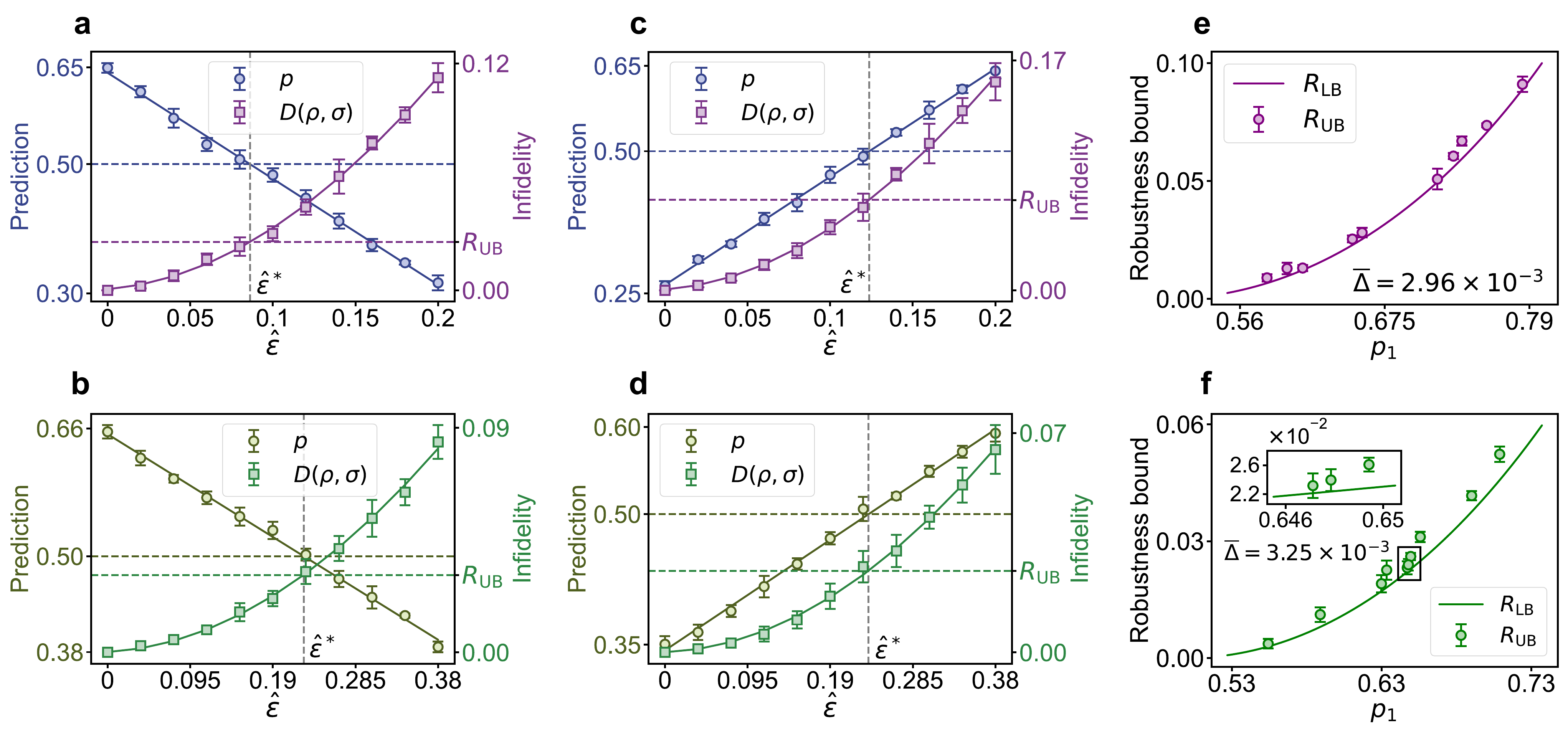}
\caption{\textbf{Experimental benchmarking of robustness bounds.} $\textbf{a}$-$\textbf{d}$, Prediction probability $p$ and infidelity $D(\rho,\sigma)$ as functions of normalized perturbation strength $\hat{\epsilon}$. Panels $\textbf{a}$ and $\textbf{c}$ show robustness upper bound $R_{\rm{UB}}$ benchmarking experiments for two EMNIST samples corresponding to handwritten letters``Q'' and ``T'', respectively. Panels $\textbf{b}$ and $\textbf{d}$ present analogous results for non-excited and excited cluster states in the LCEI task. Data points ($n=5$ independent experiments) are shown as mean $\pm$ standard deviation (SD). Solid lines represent fitting curves of $p_1$ and $D(\rho,\sigma)$, while the dashed purple/green lines mark the extracted values of $R_{\rm{UB}}$ at $\hat{\epsilon}^*$. $\textbf{e}$, $\textbf{f}$, Comparison of experimentally extracted  upper bound $R_{\rm{UB}}$ (from $10$ randomly selected samples, $5$ per class) versus the theoretical lower bound $R_{\rm{LB}}$.  Error bars indicate the root mean square error from fitting $D(\rho,\sigma)$. The averaged gap $\overline{\Delta}$ indicates the average gap between $R_{\rm{UB}}$ and $R_{\rm{LB}}$ across the $10$ samples, as defined under the infidelity metric.}
\label{fig2}
\end{figure*}

\begin{figure*}[!th]
\centering
\includegraphics[width=0.95\textwidth]{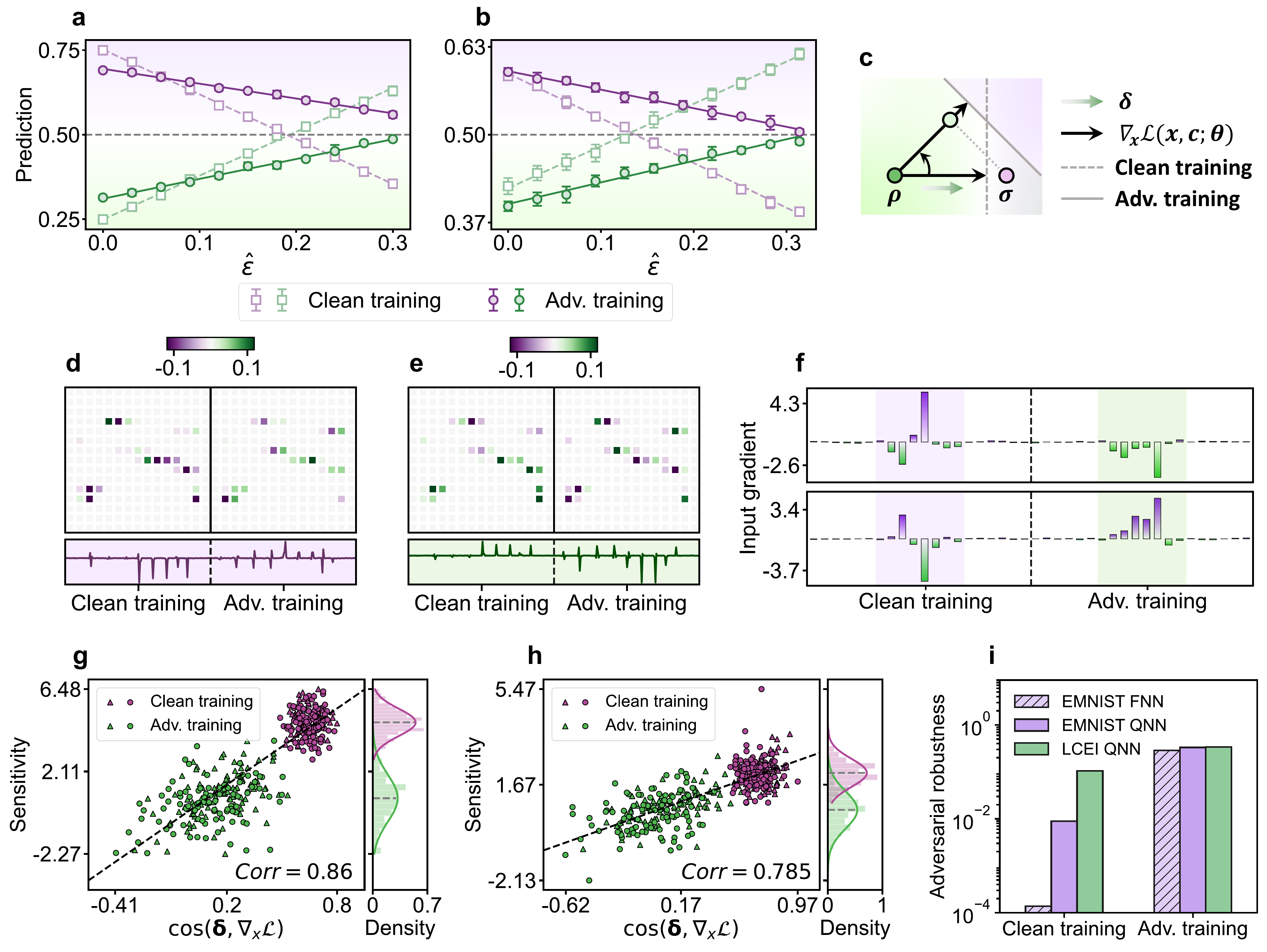}
\caption{\textbf{Adversary robustness benchmarking experiments.} $\textbf{a}$, $\textbf{b}$, QNN sensitivity experiment for EMNIST (\textbf{a}) and LCEI (\textbf{b}). Purple and green indicate distinct classes. Datas ($n=5$ independent experiments) are presented in the form of mean values $\pm$ SD, with solid and dashed lines denoting linear fits for clean and adversarial trained models, respectively. $\textbf{c}$, Diagram of the local classification landscape, illustrating how adversarial training reorients input gradients to enhance robustness. $\textbf{d}$, $\textbf{e}$, Comparison of average input gradients for images ``Q'' ($\textbf{d}$) and ``T'' ($\textbf{e}$). The top panels show per-pixel gradients, while the bottom panels show flattened gradient vectors to facilitate observation. $\textbf{f}$, Comparison of average input gradients for LCEI states. The top panel shows the gradients of rotation angle for excited clusters, while the bottom for non-excited clusters. $\textbf{g}$, $\textbf{h}$, Correlation between the sensitivity $S$ and cosine similarity $\operatorname{cos}(\boldsymbol{\delta},\nabla_{\boldsymbol{x}} \mathcal{L})$ for EMNIST ($\textbf{g}$) and LCEI ($\textbf{h}$), with triangles and circles for different classes. The right panels show the sensitivity distribution under clean and adversarial training. $\textbf{i}$, Adversarial robustness scores of QNNs across different datasets under various training strategies, along with comparisons to FNN. }
\label{fig3}  
\end{figure*}

\noindent Our research focuses on a QNN classifier based on variational quantum circuits~\cite{benedetti2019parameterized,Cerezo2021,Bharti2022}, as illustrated in Fig.~\ref{fig1}\textbf{a}. The variational parameters $\boldsymbol{\theta}$ are optimized to minimize the empirical risk $J(\boldsymbol{\theta}) =  \mathbb{E}_{(\boldsymbol{x},k) \in \mathcal{D}} \left[ \mathcal{L}(f(\boldsymbol{x};\boldsymbol{\theta}),\boldsymbol{c}) \right],$ where $\mathcal{D}$ is training set, $f(\boldsymbol{x};\boldsymbol{\theta})$ represents the hypothesis function of the QNN classifier, $\boldsymbol{c}$ is one-hot encoding of label $k$, and \textbf{$\mathcal{L}$} is the applied cross-entropy loss applied function. This work evaluates two key metrics of QNN robustness: robustness bound and adversarial robustness. Robustness bounds quantify the minimum perturbation distance required to induce misclassification (Fig.~\ref{fig1}\textbf{b}). Adversarial robustness is benchmarked by measuring the model's performance under specific perturbations: high sensitivity to attack indicates poor robustness, whereas low sensitivity implies strong robustness. (Fig.~\ref{fig1}\textbf{c}). Our experiments encompass both classical image classification tasks (handwritten letters ``Q'' and ``T'', Fig.~\ref{fig1}\textbf{d}) and quantum linear cluster excitation identification (LCEI), where the QNN is trained to distinguish between $20$-qubit cluster states in excited and non-excited states within a synthetic dataset (Fig.~\ref{fig1}\textbf{e}), see Appendix for details. In both binary classification tasks, the QNN predicts with probability $p = (\langle \sigma_z \rangle + 1)/2$ from the measured expectation of the Pauli-$z$ operator on the output qubit, with the decision boundary set at $p = 0.5$.

Accurate and efficient adversarial attack algorithms are crucial for probing a model's robustness. To this end, we propose the masked adversarial attack (Mask-FGSM), a localized attack variant of the fast gradient sign method (FGSM)~\cite{Goodfellow2014}, designed for efficient attacks and fast generation of adversarial samples in QML experiments (Fig.~\ref{fig1}\textbf{f}). The method perturbs only a strategically selected sparse subset of input features, defined by a binary mask $\mathcal{M} \in \{0,1\}^{\operatorname{dim}(\boldsymbol{x})}$. This sparse perturbation strategy is adopted to minimize the computational cost of the attack and mitigate the impact of noisy gradients on attack effectiveness, especially in experimental settings. Adversarial samples are generated by perturbing legitimate samples as
\begin{eqnarray}\label{eq:mask_FGSM}
	\boldsymbol{x}' = \boldsymbol{x} + \epsilon \cdot \left( \mathcal{M} \odot \operatorname{sign}({\nabla}_{\boldsymbol{x}} \mathcal{L}) \right).
\end{eqnarray}
where $\epsilon$ denotes the perturbation strength and $\odot$ denotes elementwise product. The binary mask is derived by analyzing the magnitude of input gradient components. A step-by-step summary of this workflow is detailed in the Appendix.

Our experiments are executed on a frequency-tunable flip-chip superconducting quantum processor with $72$ qubits and $126$ couplers in a two-dimensional array (Fig.~\ref{fig1}\textbf{g}). For the experiment, we selected a one-dimensional (1D) array of $20$ high-quality qubits, where arbitrary single-qubit gates and two-qubit controlled-Z (CZ) gates can be implemented for each qubit and adjacent qubit pairs. To achieve high-fidelity parallel gate operations required by the algorithm, we constructed an error model to optimize the operating frequencies of all qubits and the interaction frequencies of two-qubit gates, avoiding two-level system (TLS) regions and suppressing residual coupling. See Supplementary Sections VI and VII for the device parameters and calibration details.

\vspace{.5cm}
\noindent\textbf{\large{Robustness bound}}{\large\par}

\noindent Robustness bounds provide a principled measure of a classifier’s ability to maintain correct predictions under adversarial perturbations~\cite{weng2018evaluating,ko2019popqorn}. In general, for a given input, the robustness radius—the minimum distance required to induce misclassification-is NP-hard to compute exactly, as it requires solving a nonconvex optimization problem over high-dimensional perturbation directions. Therefore, two complementary strategies are typically employed: (i) analytical derivation of a provable lower bound that guarantees robustness within a certified radius, and (ii) empirically estimating an upper bound via explicitly constructing adversarial examples through effective attacks. Together, these two bounds bracket the true robustness radius and provide a practically accessible yet theoretically meaningful robustness estimation.

We adapt this framework to QNNs by employing the state infidelity as the perturbation metric. Specifically, for a legitimate quantum input state $\rho$, we define the distance to a perturbed state $\sigma$ as $D(\rho,\sigma)=1-F(\rho,\sigma)$, where $F(\rho,\sigma)=(\mathrm{Tr}\sqrt{\sqrt{\rho}\sigma\sqrt{\rho}})^2$. Within this metric space, robustness bounds of a QNN are formalized as follows:
\begin{definition}[Robustness lower bound]
	$R_{\rm{LB}}$ is a robustness lower bound of $\rho$ if there is no adversarial example exist when perturbed states $\sigma$ satisfied $D(\rho,\sigma)<R_{\rm{LB}}$.
\end{definition}
\begin{definition}[Robustness upper bound]
	$R_{\rm{UB}}$ is a robustness upper bound of $\rho$ if there exists at least one adversarial example when perturbed states $\sigma$ satisfy $D(\rho,\sigma) \geq R_{\rm{UB}}$.
\end{definition}
The closed-form analytical expression for the sample-wise robustness lower bound has been provided in Ref.~\cite{Guan2020}:
\begin{eqnarray}\label{eq: R_LB}
	R_{\rm{LB}} = \frac{1}{2}\left(\sqrt{p_1}-\sqrt{p_2}\right)^2,
\end{eqnarray}
where $p_1$ and $p_2$ denote the predicted probabilities associated with the correct and competing classes, see Supplementary Section IV for details. In the binary classification case, this simplifies to $p_1 = \operatorname{max}(p, 1-p)$, providing a computationally efficient certification of robustness for each sample.

In contrast, estimating the upper bound $R_{\rm UB}$ requires explicit construction of adversarial states. We physically generate adversarial states on the superconducting quantum processor using iterative Mask-FGSM attacks. At each perturbation step, we simultaneously measure the prediction probability $p$ and compute the infidelity $D(\rho,\sigma)$ from quantum state tomography (QST) performed on the reduced density matrix of the output qubit, both as functions of the normalized perturbation strength $\hat{\epsilon}=\epsilon/(\boldsymbol{x}_{\rm{max}} - \boldsymbol{x}_{\rm{min}})$ (Fig.~\ref{fig2}\textbf{a-d}). In gradient-based attack experiments, both $p$ and $D(\rho,\sigma)$ are well fitted by a $\operatorname{cos}^2$-type function, consistent with the sinusoidal response of single-qubit rotations under perturbations. From these fits, we extract the critical perturbation threshold $\hat{\epsilon}^*$ at the misclassification boundary ($p = 0.5$). and compute the corresponding infidelity, which yields empirical $R_{\rm{UB}}$. 

We compare the experimentally extracted $R_{\rm UB}$ against the analytical $R_{\rm LB}$ of Eq.~(\ref{eq: R_LB}). As shown in Fig.~\ref{fig2}\textbf{e}, \textbf{f} for ten randomly chosen samples, $R_{\rm{UB}}$ exceed $R_{\rm{LB}}$ by only $2.96 \times 10^{-3}$ on EMNIST and $3.25 \times 10^{-3}$ on LCEI, demonstrating that the adversarial attack strategy nearly saturates the theoretical robustness limit. This constitutes the first hardware-level experimental validation of the tightness of robustness bounds, establishing masked adversarial attacks as an operationally near-optimal probing method for quantum robustness.

\vspace{.5cm}
\noindent\textbf{\large{Adversarial robustness}}
{\large\par}

\noindent Beyond these static metrics, a comprehensive understanding of QNN resilience requires evaluating their dynamic stability to adversarial perturbations. We measured the sensitivity of QNNs under both clean and adversarial training. Adversarial training, a widely recognized defense strategy, enhances robustness by incorporating adversarial examples into the dataset. Fig.~\ref{fig3}\textbf{a}, \textbf{b} show prediction $p$ as function of $\hat{\epsilon}$. The observed change $\Delta p$ intuitively reflects adversarial robustness: clean-trained models exhibit dramatic output changes under attack, leading to mispredictions, whereas adversarial-trained models demonstrate significantly reduced sensitivity. This robustness enhancement arises from the implicit regularization of the inner product between the perturbation vector $\boldsymbol{\delta}$ and the input gradient $\nabla_{\boldsymbol{x}} \mathcal{L}$, which reorients gradients to mitigate alignment with adversarial directions (Fig.~\ref{fig3}\textbf{c}), with a detailed proof provided in Appendix. Input gradient visualizations further illustrate this mechanism (Fig.~\ref{fig3}\textbf{d}-\textbf{f}), adversarial training preserves the critical feature regions of the inputs but modifies the magnitude of the gradient components, thereby reducing their overlap with $\boldsymbol{\delta}$.

We propose a logit sensitivity-based adversarial robustness score for quantitative analysis. Sensitivity $S = \Delta L / \hat{\epsilon}$ measures the impact of input perturbations on the logit space, where $\Delta L = L(\boldsymbol{x}+\boldsymbol{\delta})-L(\boldsymbol{x})$. We adopt the logit representation since the probability space suffers from saturation effects at high confidence levels. For QNNs, we introduce a ``quantum logit'', as the quantum analogue of the unnormalized prediction score in classical models before the final nonlinear activation. It maps the $\langle \sigma_z \rangle$ onto the real space $z \in \mathbb{R}$:
\begin{eqnarray}\label{eq: qulogit}
	L = \operatorname{ln} \left(\frac{p }{1-p } \right) = \operatorname{ln} \left( \frac{1 + \langle \sigma_z \rangle }{1 - \langle \sigma_z \rangle } \right). 
\end{eqnarray}
Then, the adversarial robustness score of a QNN is defined as the average robustness across samples in the dataset, expressed as a mirrored sigmoid function of sensitivity:
\begin{eqnarray}\label{eq:R_loc}
	\overline{R}_{\rm{adv}} = \frac{1}{|\mathcal{D}|} \sum_{\boldsymbol{x} \in \mathcal{D}} \frac{1}{1+e^{S(\boldsymbol{x})}}.
\end{eqnarray}
To quantify the change in the QNN’s sensitivity to specific adversarial attacks, we conducted a statistical experiment involving $200$ samples ($100$ from each class). Fig.~\ref{fig3}\textbf{g}, \textbf{h} present scatter plots of $S$ versus cosine similarity $\operatorname{cos}(\boldsymbol{\delta}, \nabla_{\boldsymbol{x}} \mathcal{L})$ for each sample in EMNIST and LCEI. The results confirm that adversarial training reduces QNN sensitivity to target perturbations. Moreover, strong correlations were observed between $S$ and $\operatorname{cos}(\boldsymbol{\delta}, \nabla_{\boldsymbol{x}} \mathcal{L})$, with Pearson correlation coefficients of $0.86$ (EMNIST) and $0.785$ (LCEI), providing substantial experimental support for the theoretical framework of gradient regularization.

We computed $\overline{R}_{\rm{adv}}$ across these $200$ samples, with results shown in Fig.~\ref{fig3}\textbf{i}. Under clean training, QNNs exhibit higher robustness on quantum datasets than on classical datasets, reflecting fundamental differences in feature space structure. Image feature spaces are notoriously laden with adversarial traps. These traps stem from mathematically proximate yet semantically distinct features, which attackers can exploit by shifting along high-dimensional, unconstrained directions to deceive the model easily. By contrast, feature states used in quantum datasets are typically endowed with specific physical significance, where perturbing a meaningful configuration (such as an unexcited cluster) to a different class requires a costly physical operation. Extending this analysis, in the largest-scale adversarial learning experiment, we showed that adversarial training increased the QNN robustness score from $8.90 \times 10^{-3}$ to $0.330$ on EMNIST, and from $0.105$ to $0.339$ on LCEI. These results demonstrate that our metric faithfully captures robustness differences across datasets and training strategies.

With this metric, we also compared the adversarial robustness of QNNs with classical feedforward neural networks (FNNs), ensuring comparable network sizes, optimizers, and accuracies on the same EMNIST dataset. We find that QNNs achieve significantly higher adversarial robustness scores than FNNs, with clean-trained QNNs outperforming FNNs by over $64$-fold. We attribute this experimentally observed robustness enhancement to noise-induced gradient attenuation. The inherent hardware noise reduces the model's sensitivity to all input variations, effectively masking the fine-grained perturbations used in attacks while preserving the dominant features required for classification. Specifically, we model NISQ noise on the output qubit as a depolarizing channel with contraction factor $\xi \propto t/T_1 + t/T_2$ (where $T_1$ and $T_2$ are relaxation and dephasing times), approximated via twirling under a unitary 2-design. After $l$ circuit layers, this contraction scales the $\Delta L$ by $(1-\xi)^l$, yielding noisy sensitivity 
\begin{eqnarray}\label{eq:xi}
    S^{\rm{noisy}}= \underbrace{\frac{ (1-\xi)^l \left( 1-z_0^2 \right)}{1-(1-\xi)^{2l}   z_0^2}}_{:=C(\xi,l,z_0)} S^{\rm{ideal}},
\end{eqnarray}
where $z_0=\langle \sigma_z \rangle$ in noiseless case. We prove $C<1$ for $\xi > 0$, indicating that noise reduces sensitivity and thereby increases the robustness score (see Supplementary Section V for details). We note that, while classical methods have attempted to improve robustness through noise injection~\cite{he2019parametric,liu2021training,ye2023improving}. Our results suggest that the intrinsic noise in NISQ devices can act as a natural gradient mask, providing a distinct mechanism for resisting adversarial attacks compared to classical models.

\vspace{.5cm}
\noindent\textbf{\large{}Conclusion}{\large\par}

\noindent In summary, this work presents the first experimental benchmark of robustness for $20$-qubit QNN on a superconducting quantum processor, establishing a comprehensive framework to characterize both vulnerabilities and defense strategies. The proposed Mask-FGSM effectively identifies fragile subspaces of QNNs as a scalable and experimentally accessible diagnostic tool for QML, providing a general experimental methodology for future robustness evaluation. Analysis based on robustness score metrics demonstrated that adversarial training substantially enhances robustness against targeted attacks. Comparative experiments further demonstrate that QNN exhibits stronger robustness than classical FNN, suggesting that noisy quantum hardware can act as a form of gradient masking, providing distinct security characteristics compared to noiseless models.

Despite these advances, several challenges remain. Adversarial training, while effective at improving task-specific robustness, cannot provide universal protection, as adversarial examples are ubiquitous in high-dimensional quantum feature spaces~\cite{Liu2020,gong2022universal}. Quantum adversarial learning thus acts primarily as a patching mechanism against known attacks, while absolute robustness is unattainable for non-trivial classifiers since enforcing it would collapse the model into a constant function. This limitation parallels insights from classical machine learning, where techniques such as defensive distillation~\cite{2016distillation} and gradient regularization~\cite{ross2018} have been proposed, but their quantum analogs remain largely unexplored. Future research should extend the robustness evaluation framework to a broader class of QMLs, including unsupervised models such as variational quantum autoencoders. Further exploration of advanced defense strategies, while extending the robustness estimation to large-scale fault-tolerant quantum computing, is an essential pathway toward achieving reliable quantum AI in high-risk real-world domains.

\vspace{.5cm}
\noindent\textbf{\large{}Method}{\large\par}
\noindent\textbf{Training QNN classifiers}

\begin{figure}[!t]
	\centering
	\includegraphics[width=0.25\textwidth]{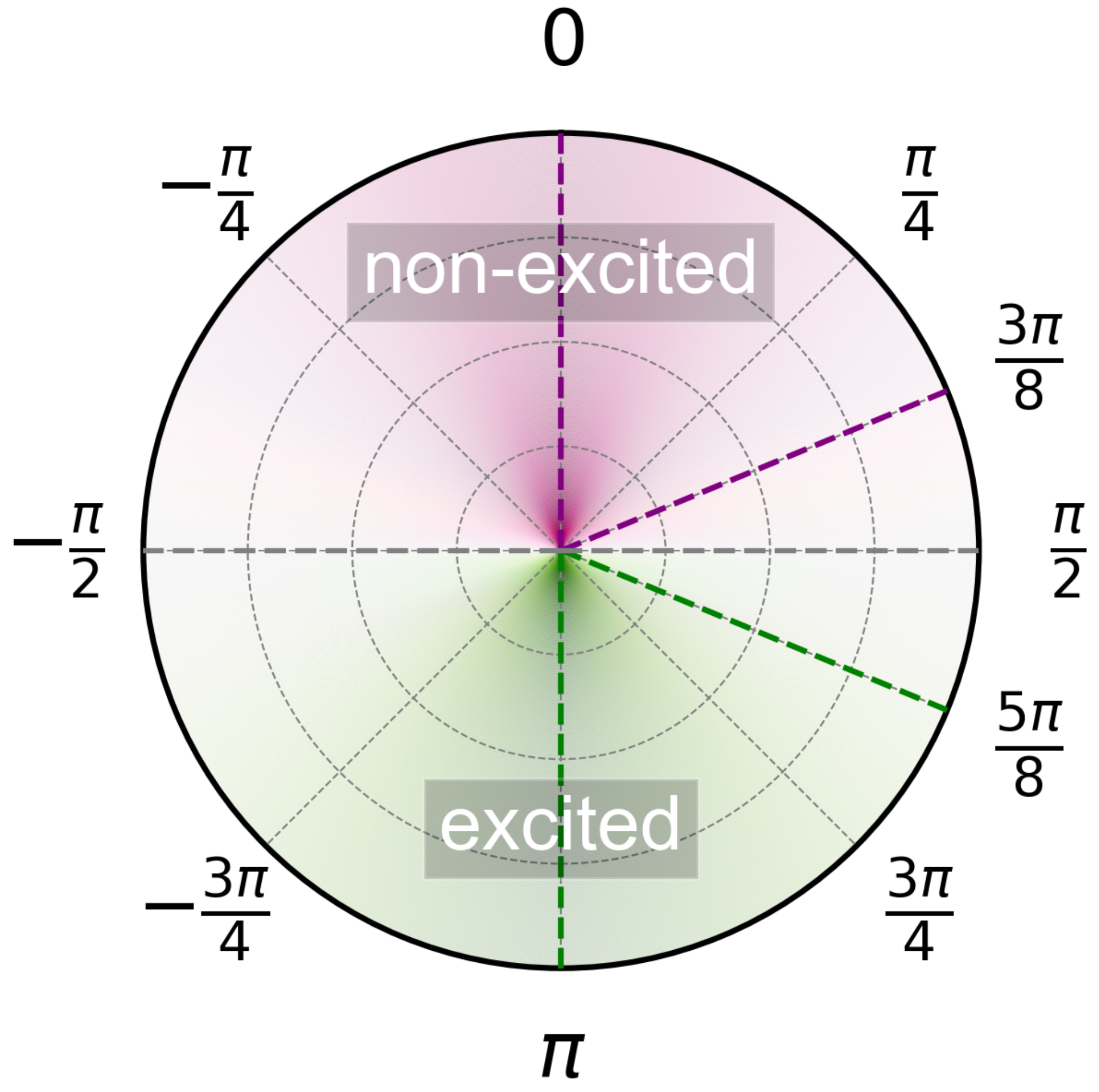}
	\caption{\textbf{Distribution of $\alpha$ for the synthetic quantum dataset in LCEI.} Quantum data for the two classes are sampled from $\alpha \in [0,{3\pi}/{8}]$ (purple dashed lines) and $[{5\pi}/{8},\pi]$ (green dashed lines).}
	\label{fig4}  
\end{figure}

\begin{figure*}[t]
	\centering
	\includegraphics[width=0.9\textwidth]{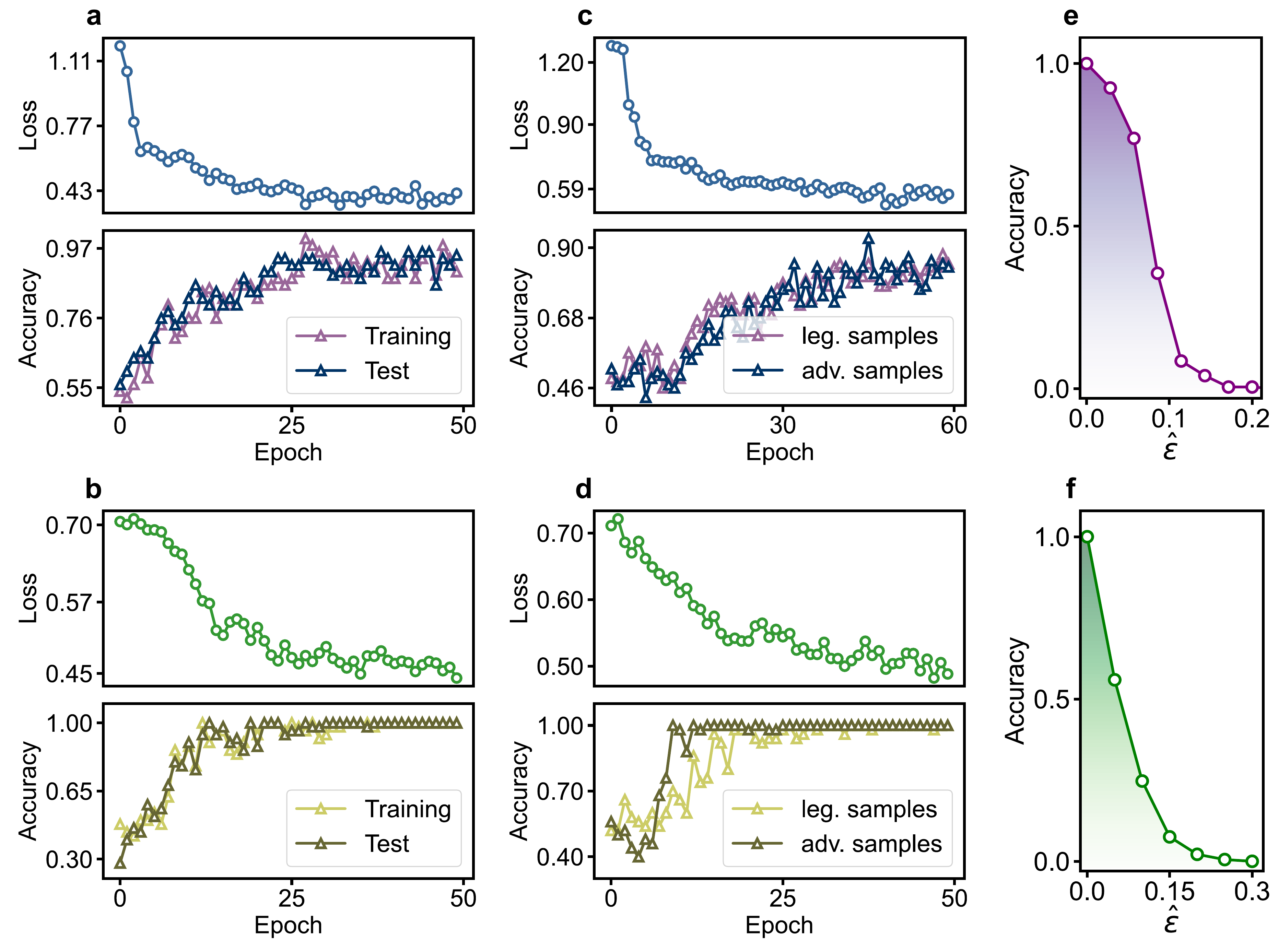}
	\caption{\textbf{Training and attack results.} $\textbf{a}$, $\textbf{b}$, Clean training: loss, training accuracy, and test accuracy as functions of epoch. $\textbf{c}$, $\textbf{d}$, Adversarial training: loss, accuracy on legitimate samples, and accuracy on adversarial samples as functions of epoch. $\textbf{e}$, $\textbf{f}$, Accuracy of perturbed samples under Mask FGSM attacks, plotted as functions of normalized perturbation strength. Panels $\textbf{a}$, $\textbf{c}$ and $\textbf{e}$ correspond to EMNIST dataset, while $\textbf{b}$, $\textbf{d}$ and $\textbf{f}$ to LCEI.}
	\label{fig5}  
\end{figure*}

\noindent Here, we detail the implementation of the two QNN classifiers investigated in this study, covering three key aspects: dataset, variational circuit, and training algorithm.

\textit{Dataset}. For the EMNIST task, we utilized a subset of $600$ handwritten letter images (``Q'' and ``T'', $300$ each) from the EMNIST dataset, split into $500$ for training and $100$ for testing. Images were resized to $15 \times 15$ resolution to balance encoding cost and classification accuracy. For LCEI, we generate a linear cluster state:
\begin{eqnarray}\label{eq:LC}
	|{\operatorname{LC}_n}\rangle = \left( \prod_{i=1}^{n-1}\operatorname{CZ}_{i,i+1} \right) {|{+} \rangle}^{\otimes n}.
\end{eqnarray}
Then, an $R_x(\alpha)$ gate is applied post-cluster state preparation, rotating qubits about the Bloch sphere’s $x$-axis with $-\pi \leq \alpha \leq \pi$. A sufficiently large rotation $|\alpha| \in (\pi / 2,\pi]$ defines an excited cluster state, while $|\alpha| \in [0,\pi/2)$ denotes a non-excited state. We created a synthetic quantum dataset by uniformly sampling $150$ $\alpha$ values from $[0,{3\pi}/{8}]$ and $[{5\pi}/{8},\pi]$ for the ``non-excited'' and ``excited'' classes, respectively (Fig.~\ref{fig4}). From these, $200$ states are randomly assigned to the training set, with $100$ reserved for testing.

\textit{Variational circuit}. Our QNN is based on a hardware-efficient variational circuit, comprising a multilayer structure with SU(2) gates on each qubit and nearest-neighbour CZ gates. The SU(2) gates serve as both data encoding gates and variational parameters, while native CZ gates provide entanglement within layers to reduce compilation overhead. The variational circuit used in this work consists of $40$-layer gate and $181$ trainable parameters. For the high-dimensional EMNIST image data, we apply an interleaved block encoding strategy to enhance performance on image datasets~\cite{Ren2022,Quantum2020,Haug2023}, where variational parameters and features are encoded into parameterized gates via linear combinations. For LCEI, the same variational circuit is used, with an additional cluster state preparation stage at the front for dataset encoding, as detailed in Supplementary Section I.

\textit{Training algorithm}. To optimize training efficiency and classification accuracy, we integrate mini-batch stochastic gradient descent with the Adam optimizer~\cite{kingma2014adam} for the training of the $20$-qubit QNN classifier. Details of the training procedure are provided in the Supplementary Section II. Clean training results (Fig.~\ref{fig5}\textbf{a}, \textbf{d}) show optimal EMNIST accuracies of 
$98\%$ (training) and $96\%$ (test), and $100\%$ for both in LCEI. Training accuracy reflects the classifier’s performance on mini-batches per epoch, while  test accuracy denotes its performance on the test set.

\vspace{.3cm}
\noindent\textbf{Attack Algorithm}

\noindent To expedite the batch generation of high-dimensional adversarial examples on quantum hardware, we propose and implement a localized attack strategy (Mask-FGSM) leveraging input-gradient sparsity. This protocol identifies the input coordinates with the greatest impact on model output, confining adversarial perturbations to these coordinates. Here, we provide an overview of the procedure (see Supplementary Section III for details).

1. \textbf{Gradient Estimation}. Compute per-feature input gradients $\nabla_{\boldsymbol{x}} \mathcal{L}$ on device for a small test mini-batch of $T$ representative samples. Form the averaged absolute-gradient vector
    \begin{eqnarray}
    	\overline{\mathcal{G}}= \frac{1}{T}\sum_{i=1}^{T}|\nabla_{\boldsymbol{x}^{(i)}} \mathcal{L}| \in \mathbb{R}^{\operatorname{dim}(\boldsymbol{x})}.
    \end{eqnarray}
    where $|\cdot|$denotes componentwise absolute value and $N_x=\operatorname{dim}(\boldsymbol{x})$.

2. \textbf{Mask Construction}. The components of $\overline{\mathcal{G}}$ are sorted in descending order of magnitude, and the index set $\mathcal{I}_{\rm{top}}$, comprising the top $r N_x$ components, is selected, from which a binary mask $\mathcal{M}$ is constructed:
    \begin{eqnarray}
	\mathcal{M}_i = \left\{
	\begin{aligned}
		&1, &&i \in \mathcal{I}_{\rm{top}}, \\
		&0, && otherwise.\\
	\end{aligned}
	\right.
\end{eqnarray}
    
3. \textbf{Attack Execution}. The FGSM or other gradient-based attacks are constrained to the support defined by the mask, as outlined in Eq.~(\ref{eq:mask_FGSM}) of the main text, with the same fixed mask reused across subsequent batch attacks.

This attack strategy computes gradients solely for mask-selected coordinates, reducing the overall gradient estimation and circuit execution frequency by approximately a factor of $1/r$. Moreover, the mask eliminates minor, noise-dominated coordinates, enhancing the reproducibility of adversarial examples under limited-shot conditions. The same procedure applies to pixel space (classical data) and parameter space (quantum data), facilitating scalable, batch-processed robustness benchmarking on NISQ platforms.

In specific experiments, for the EMNIST, we compute $\overline{\mathcal{G}}$ across $20$ test samples, empirically selecting $r \approx 0.15$ (activating the top $15\%$ of pixels), with this mask encompassing $\gtrsim 90\%$ of the $L_1$ gradient norm and applied to all Mask-FGSM batches. For LCEI, we apply an analogous procedure to the parameter space, where gradients target each $R_x$ rotation angle $\alpha$. Measurements reveal gradient concentration on qubits $\rm{Q7}$ to $\rm{Q14}$, and the corresponding angle mask is used to constrain attacks within the parameter space.

\vspace{.3cm}
\noindent\textbf{Adversarial training}

\noindent We employed the masked attack to generate $200$ adversarial samples ($100$ per class) for EMNIST and LCEI, which are subsequently used for adversarial training. Specifically, in each training epoch, mini-batches comprise $50\%$ legitimate samples randomly selected from the original dataset and $50\%$ adversarial samples from the generated set. The adversarial training process is depicted in Fig.~\ref{fig5}\textbf{b}, \textbf{e}. Following adversarial training, accuracies on the EMNIST dataset for legitimate and adversarial samples are $88\%$ and $87\%$, respectively, while LCEI achieves $100\%$ for both. To understand how adversarial training enhances robustness, we express the empirical risk in adversarial training as:
\begin{eqnarray}\label{eq:risk_adv1}
		\tilde{J}(\boldsymbol{\theta}) = \mathbb{E}_{(\boldsymbol{x},k) \in \mathcal{D}} 
		\left[ \frac{1}{2}\mathcal{L}(\boldsymbol{x},\boldsymbol{c} ;\boldsymbol{\theta}) + \frac{1}{2}\mathcal{L}(\boldsymbol{x}+\boldsymbol{\delta},\boldsymbol{c} ;\boldsymbol{\theta}) \right].
\end{eqnarray}

For small perturbations $\boldsymbol{\delta}$, we can perform the first-order Taylor expansion of $\mathcal{L}(\boldsymbol{x}+\boldsymbol{\delta},\boldsymbol{c} ;\boldsymbol{\theta})$ as:
	\begin{eqnarray}\label{eq:risk_adv2}
	\begin{aligned}
		\mathcal{L}(\boldsymbol{x}+\boldsymbol{\delta},\boldsymbol{c};\boldsymbol{\theta}) =  \mathcal{L} &(\boldsymbol{x},\boldsymbol{c};\boldsymbol{\theta}) + \nabla_{\boldsymbol{x}}\mathcal{L}(\boldsymbol{x},\boldsymbol{c};\boldsymbol{\theta}) \cdot \boldsymbol{\delta} \\ + 
		& O(\boldsymbol{\delta}^2).
	\end{aligned}
\end{eqnarray}
Therefore, 
\begin{eqnarray}\label{eq:risk_adv3}
	\begin{aligned}
		\tilde{J}(\boldsymbol{\theta}) &\approx  \mathbb{E}_{(\boldsymbol{x},k) \in \mathcal{D}} \left[ \mathcal{L}(\boldsymbol{x},\boldsymbol{c} ;\boldsymbol{\theta}),\boldsymbol{c}) + \mathcal{R} \right], \\
		\mathcal{R} &= \frac{1}{2} \nabla_{\boldsymbol{x}}\mathcal{L}(\boldsymbol{x},\boldsymbol{c} ;\boldsymbol{\theta}) \cdot \boldsymbol{\delta}.
	\end{aligned}
\end{eqnarray}

That is, adversarial training introduces a regularization term $\mathcal{R}$ to the loss function, which is the inner product of the input gradient and perturbation. Minimizing this term during training encourages turning the direction of the input gradient away from the target perturbation, thereby reducing the model’s sensitivity and enhancing adversarial robustness.

\vspace{.3cm}
\noindent\textbf{Classical neural network for comparison}

\noindent To investigate the potential robustness advantage in QML, we designed a three-layer fully connected FNN to compare with QNN. We utilized the same EMNIST dataset, with images compressed to $9 \times 9$ pixels prior to input into the FNN. The FNN consists of an input layer with $81$ neurons to encode the images, a hidden layer with $2$ neurons, and an output layer with a single neuron, totaling $167$ trainable parameters (weights and biases), comparable in scale to the QNN in the main text. ReLU activation functions are applied to the input and hidden layers to provide nonlinearity, while a sigmoid activation function maps the output to the interval $[0, 1]$ for binary classifications. The FNN was trained using the same loss function and optimizer as the QNN until achieving equivalent classification accuracy. We evaluated the adversarial robustness of the well-trained QNN and FNN on the same EMNIST dataset. Adversarial training followed identical strategies for both models.

To further validate that the observed robustness advantage of the QNN is not an artifact of the baseline FNN being under-parameterized, we performed a systematic numerical simulation comparing QNNs and FNNs with matched parameter counts across increasing scales.

Since the structural differences between quantum and classical networks prevent exact parameter equivalence, we designed a scaling protocol to maintain approximately equal parameter counts. 
For the QNN, we utilized a $3$-qubit re-uploading architecture~\cite{Quantum2020} where each layer consists of $9$ parameterized gates to encode the $N_x=81$ dimensional input, yielding $2N_x=162$ parameters per logical block. The total parameters scale as $N_{\rm{QNN}} = 2 N_x \cdot N_r$, where $N_r$ is the number of re-uploading layers.
For the FNN, we fixed the input layer at $N_x=81$ and output at $1$, varying the hidden layer size $N_h$. The total parameters scale as $N_{\rm{FNN}} = N_x \cdot N_h + 2N_h + 1$. By setting $N_r$ from $1$ to $10$ and adjusting $N_h$ accordingly, we swept the parameter space from $\sim 160$ to $\sim 1600$. The simulation was conducted under ideal, noise-free conditions with identical hyperparameters and initialization schemes. The comparative adversarial robustness scores are summarized in Table~\ref{tab:sim_comparison}.

\begin{table}[h]
\caption{\label{tab:sim_comparison} 
Comparison of $\overline{R}_{\rm{adv}}$ between QNN and FNN in noiseless simulation with matched parameter counts.}
\begin{ruledtabular}
\centering
\begin{tabular}{ccccc}
Scale ($N_r$) & $N_{\rm{QNN}}$ & $N_{\rm{FNN}}$ & $\overline{R}_{\rm{QNN}}$ (Sim.) & $\overline{R}_{\rm{FNN}}$ (Sim.) \\
\hline
1  & 162  & 167  & 0.105 & 0.075 \\
2  & 324  & 333  & 0.104 & 0.079 \\
3  & 486  & 499  & 0.077 & 0.076 \\
4  & 648  & 665  & 0.082 & 0.083 \\
5  & 810  & 822  & 0.075 & 0.089 \\
6  & 972  & 985  & 0.040 & 0.079 \\
7  & 1134 & 1149 & 0.071 & 0.063 \\
8  & 1296 & 1313 & 0.077 & 0.083 \\
9  & 1458 & 1477 & 0.085 & 0.103 \\
10 & 1620 & 1661 & 0.088 & 0.062 \\
\end{tabular}
\end{ruledtabular}
\end{table}

The results reveal no consistent robustness advantage for the QNN over the FNN in the noiseless regime, with both models exhibiting fluctuating scores around $\sim 0.07$ regardless of scale. These noiseless simulation scores are significantly lower than the experimentally measured robustness scores of the QNN (which reached $\sim 0.33$ under adversarial training). This contrast provides compelling evidence that the superior robustness observed in our main experiment is not due to the limited capacity of the classical baseline, but is indeed driven by the intrinsic quantum noise in the NISQ processor. As discussed in Supplementary Materials Sec. VI, this noise induces gradient attenuation, which significantly lowers the adversarial sensitivity compared to the noiseless FNN baseline.


\vspace{.5cm}
\noindent\textbf{\large{}Conflict of interest}{\large\par}
\noindent The authors declare that they have no conflict of interest.

\vspace{.5cm}
\noindent\textbf{\large{Acknowledgements}}{\large\par}
\noindent This work has been supported by the National Key Research and Development Program of China (Grant No. 2023YFB4502500), the National Natural Science Foundation of China (Grant No. 12404564), and the Anhui Province Science and Technology Innovation (Grant No. 202423s06050001). This work is partially carried out at the USTC Center for Micro and Nanoscale Research and Fabrication. 

\let\oldaddcontentsline\addcontentsline
\renewcommand{\addcontentsline}[3]{}

\normalem
\bibliography{ref1}

\let\addcontentsline\oldaddcontentsline

\clearpage

\newcommand{\beginsupplement}{%
	\setcounter{table}{0}%
	\renewcommand{\thetable}{S\arabic{table}}%
	\setcounter{figure}{0}%
	\renewcommand{\thefigure}{S\arabic{figure}}%
	\setcounter{equation}{0}%
	\renewcommand{\theequation}{S\arabic{equation}}%
	\setcounter{section}{0}%
	\renewcommand{\thesection}{\Roman{section}}%
	\setcounter{page}{1}%
	\renewcommand{\thepage}{S\arabic{page}}%
}


\onecolumngrid

\begin{center}
	\textbf{\large Supplementary materials for \\ ``Experimental robustness benchmarking of quantum neural networks on a superconducting quantum processor''}
\end{center}

\maketitle
\tableofcontents
\beginsupplement

\section{Quantum neural network classifier}

 The quantum neural network (QNN) classifier $\mathcal{A}$ is defined by a variational quantum circuit $U(\boldsymbol{x},\boldsymbol{\theta})$ together with a measurement set $\Pi={E_i^\dagger E_i}$, satisfying $\sum_i \Pi_i = 1$. The trainable parameters $\boldsymbol{\theta}$ are optimized by a classical routine to learn a dataset $\mathcal{D}={(\boldsymbol{x},k)_i}$ encoded into the circuit. The variational circuit is structured as a depth-$l$ unitary ansatz, $U(\boldsymbol{\theta}) = \prod_{i=1}^l W_{\rm loc}^{(i)}(\boldsymbol{\theta}_i), V_{\rm ent}$, where each layer consists of local single-qubit rotations $W_{\rm{loc}}^{{i}}(\boldsymbol{\theta}_i)={\bigotimes}_jR(\boldsymbol{\theta}_{i,j})$  followed by a global entangling operation $V_{\rm ent}$. In this work, local rotations are realized by SU(2) gates, while entanglement is generated by layers of parallel controlled-$Z$ (CZ) gates. The SU(2) gate provides three degrees of freedom corresponding to arbitrary rotations on the Bloch sphere, and can be decomposed into a sequence of hardware-native operations as
 
 \begin{eqnarray}
 	U(\theta,\phi,\lambda)=R_z(\phi-\pi/2)R_x(\pi/2)R_z(\pi-\theta)R_x(\pi/2)R_z(\lambda-\pi/2),
 \end{eqnarray}
 
where $R_x$ ($R_z$) denotes rotation about the $x$- ($z$-) axis on the Bloch sphere. The decomposition leverages the virtual-$Z$ technique, enabling error-free implementation of arbitrary $R_z$ rotations~\cite{mckay2017efficient}. The CZ gate, a standard entangling primitive on superconducting qubits, is engineered by dynamically tuning the qubit frequency to bring the $|11\rangle$ and $|20\rangle$ states into resonance, with the calibration process detailed separately.

For classical datasets such as images, data must first be embedded into quantum states. This embedding corresponds to a mapping from a $d$-dimensional real vector to a $2^n$-dimensional quantum state in the Hilbert space $\mathcal{H}$ of $n$ qubits, $\mathbb{R}^d \rightarrow \mathbb{C}^{2^n}$. Conventional encoding strategies—basis, angle, and amplitude encoding—offer distinct trade-offs, but become prohibitively costly for high-dimensional images as they demand deep circuits incompatible with the limited coherence times of noisy intermediate-scale quantum (NISQ) devices. To overcome this limitation, we employ data re-uploading, also referred to as interleaved encoding, which integrates data encoding directly into the variational layers of the circuit~\cite{Ren2022,Quantum2020,Haug2023}:

\begin{figure*}[!ht]
	\centering
	\includegraphics[width=0.5\textwidth]{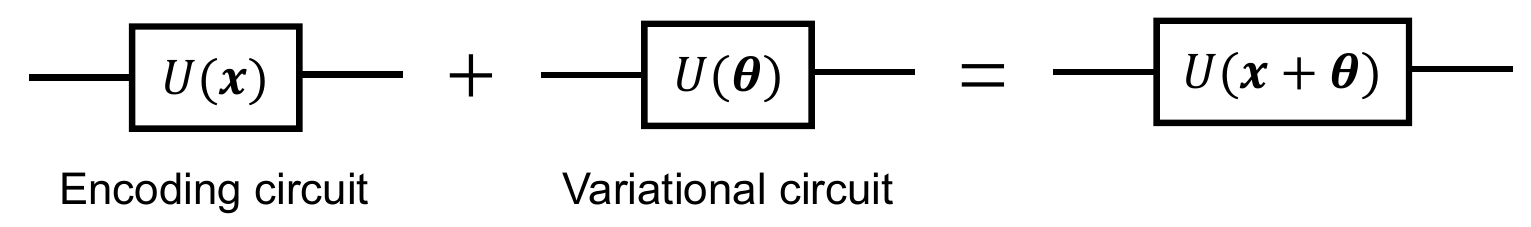}
\end{figure*}

\newpage

\begin{figure*}[!t]
	\centering
	\includegraphics[width=1\textwidth]{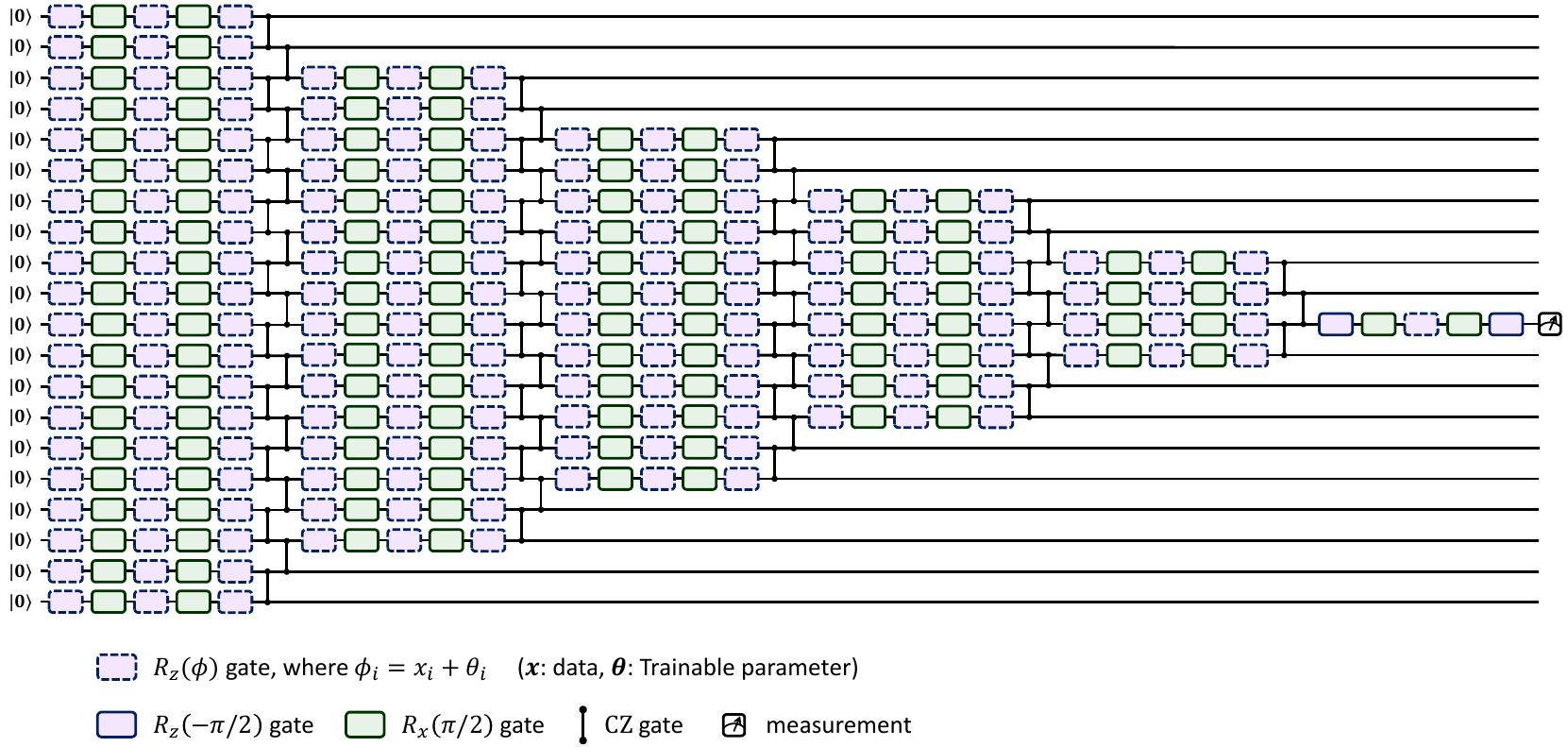}
	\caption{\textbf{Experimental QNN circuit for classification of EMNIST dataset.} The gates with dashed violet boxes of the circuit are used both for image encoding and as a training parameter. For the $i$-th parametrized $R_z$ gate, $\phi_i=x_i+\theta_i$, where $x_i$ is the $i$-th parameter of the input encoding, and $\theta_i$ is the $i$-th trainable parameter.
	}
	\label{fig:circuit_emnist}
\end{figure*}

\begin{figure*}[!ht]
	\centering
	\includegraphics[width=0.3\textwidth]{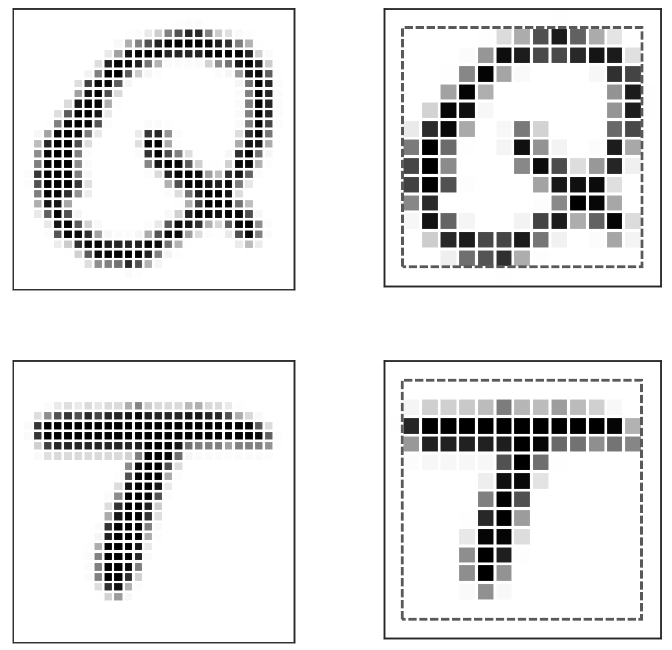}
	\caption{\textbf{Illustration of the image preprocessing and encoding strategy for handwritten letters.} The original image (left) is downscaled to $15 \times 15$ pixels (right), and then the valid region in the dashed box is extracted and encoded into the QNN circuit.
	}
	\label{fig:emnist_img}
\end{figure*}

In the case of EMNIST image classification, the dedicated QNN circuit is shown in Fig.~\ref{fig:circuit_emnist}. It comprises five blocks of decreasing size followed by a final $R_x$ rotation before measurement, amounting to $181$ trainable parameters that jointly serve as variational and angle-encoding variables. The grayscale images are downsampled via quadratic interpolation to $15\times 15$, with the central $169$ pixels encoded as rotation angles (Fig.~\ref{fig:emnist_img}). The remaining $12$ parameters are variational only and are not used for encoding. Classification is performed by measuring the central qubit: the output label is assigned according to whether the prediction probability $p$ exceeds $0.5$.

For the quantum cluster-state excitation identification (LCEI) task, we employ the same variational QNN architecture. In this case, as no classical-to-quantum encoding is required, all $181$ circuit parameters are purely variational. Instead, input states from the target dataset are prepared by a dedicated state-preparation circuit preceding the variational layers.

\begin{figure*}[!t]
	\centering
	\includegraphics[width=1\textwidth]{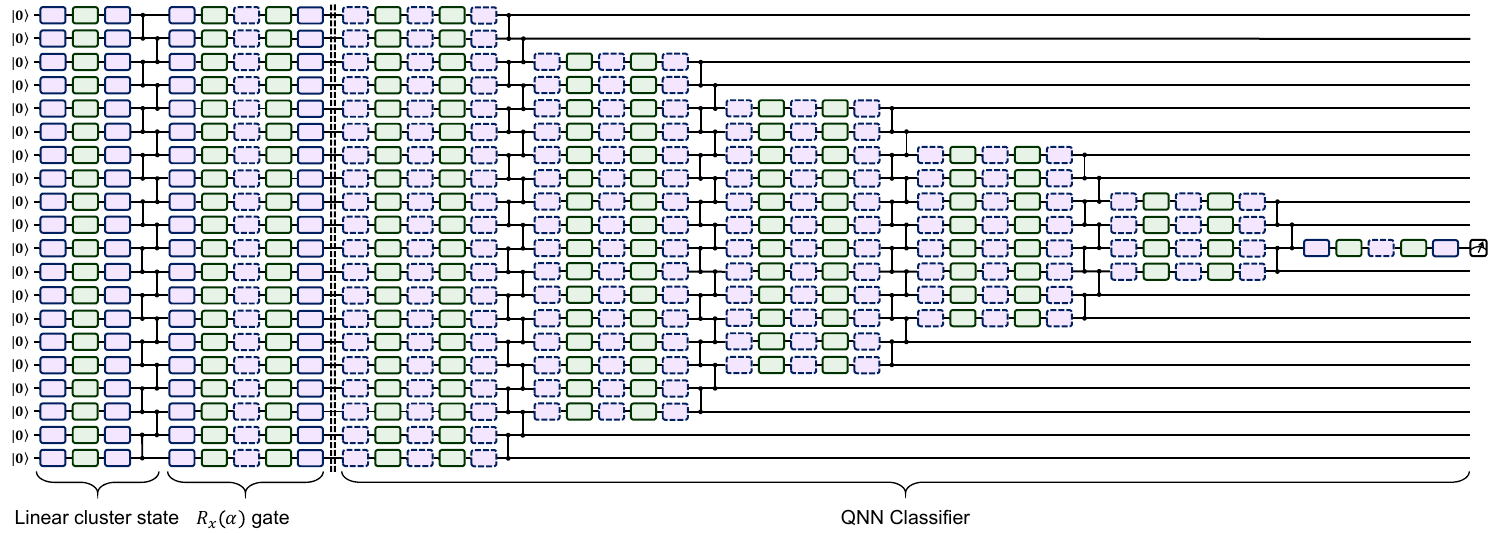}
	\caption{\textbf{Experimental QNN circuit for LCEI.} The overall circuit scheme shares the same structure as the EMNIST circuit, with a few additional layers appended at the beginning for preparing the corresponding linear cluster state.
	}
	\label{fig:LCEI}	
\end{figure*}

\section{Training algorithm}\label{sec:SI_training_algorithm}

Training a quantum classifier entails minimizing the empirical risk over a labelled dataset $\{(\boldsymbol{x}_i,\boldsymbol{c}_i)\}_{i=1}^N$. Concretely, we seek
\begin{eqnarray}\label{eq:cost}
	\underset {\boldsymbol{\theta}} { \operatorname {min} }. \  \frac{1}{N} \sum_{i=1}^{N} \sum_{k=1}^{K} \mathcal{L}({f}(\boldsymbol{x};\boldsymbol{\theta}),\boldsymbol{c}),
\end{eqnarray}
 where $\boldsymbol{\theta}$ is the parameter vector of the variational circuit to be optimized, and $\mathcal{L}$ is the loss function comparing the model output with the target label. For classification tasks, common choices include mean square error (MSE) and cross-entropy (CE) loss, with this work adopting the cross-entropy loss:
\begin{eqnarray}
	\mathcal{L}_{CE}({f}(\boldsymbol{x};\boldsymbol{\theta}),\boldsymbol{c}) = \sum_{k=1}^{K} -c_k \operatorname{log}(y_k).
\end{eqnarray}
with $y_k$ the model’s predicted probability for class $k$ and $\boldsymbol{c}$ the one-hot label vector.

A typical approach for cost minimization in Eq.~(\ref{eq:cost}) is the vanilla gradient descent (GD) method. In GD, parameters are updated iteratively based on the gradient of the loss to reduce the risk. At the $t$-th iteration, parameter ${\theta}_i$ is updates as:
\begin{eqnarray}\label{eq:GD}
	\theta_i^{(t)} = \theta_i^{(t-1)} - \eta \frac{\partial \mathcal{L} (\boldsymbol{\theta}^{(t-1)}) }{\partial \theta_i}
\end{eqnarray}
where $\eta$ is the learning rate. In variational quantum models, the network output is expressed as an expectation value ${\langle \mathcal{O}_k \rangle}_{\boldsymbol{\theta}} $, so gradients of the loss are obtained by the chain rule from derivatives ${\partial {\langle \mathcal{O}_k \rangle}_{\boldsymbol{\theta}}}/{\partial \theta_i}$. For parameterized gates of the form $U(\theta_i)=\operatorname{exp}(-i\frac{\theta_i}{2}P_i)$, where $P_i=\{I,X,Y,Z\}^{\otimes n}$ as a tensor product of Pauli operators, these partial derivatives admit the analytic parameter-shift rule (PSR)~\cite{Mitarai2018,schuld2019evaluating}:
\begin{eqnarray}\label{eq:S17}
	\frac{\partial {\langle \mathcal{O} \rangle}_{\boldsymbol{\theta}}}{\partial \theta_i} = \frac{{\langle \mathcal{O} \rangle}_{\boldsymbol{\theta} + \frac{\pi}{2} \boldsymbol{e}_i}
		-{\langle \mathcal{O} \rangle}_{\boldsymbol{\theta} - \frac{\pi}{2} \boldsymbol{e}_i}}
	{2},
\end{eqnarray}
where the subscript ${\boldsymbol{\theta} \pm \frac{\pi}{2} \boldsymbol{e}_i}$ denotes a parameter vector with the $i$-th component shifted by $\pm \pi/2$. The PSR permits unbiased gradient estimates on hardware by executing the circuit twice per parameter shift and is directly applicable to our ansatz since variational parameters are implemented as single-qubit Pauli rotations.

Computational cost considerations are central in QML. Vanilla GD computes the gradient over the entire training dataset at each iteration. It update requires evaluating gradients for all $N_p$ parameters across all $N_d$ samples with $N_s$ shots per circuit evaluation, yielding an execution complexity on the order of $O(2N_pN_dN_s)$ circuit runs per update (factor 2 from the two PSR evaluations). This becomes computationally prohibitive for large datasets. To speed up the training process, stochastic gradient descent (SGD) or mini-batch stochastic gradient descent (MB-SGD) is typically employed. SDG uses a single sample to evaluate the loss and compute the gradient, while MB-SGD, offering a better balance, approximates the loss gradient by randomly sampling a mini-batch subset $B$ (with $B \ll N_d$) from the training set during each iteration. The gradient approximation and parameter update are given by:
\begin{eqnarray}
	\boldsymbol{g} &\approx& \frac{1}{|B|} \sum_{(\boldsymbol{x}_i,\boldsymbol{c}_i) \in B} {\nabla}_{\boldsymbol{\theta}} \mathcal{L}({f}(\boldsymbol{x}_i;\boldsymbol{\theta}),\boldsymbol{c}_i), \\
	\boldsymbol{\theta}_{t} &=& \boldsymbol{\theta}_{t-1} - \eta \boldsymbol{g}_{t}.
\end{eqnarray}
Notably, while MB-SGD does not provide unbiased estimates of the gradient at each step, it substantially reduces per-iteration resource demands and empirically performs well in noisy optimization regimes typical of NISQ hardware.

Although the vanilla form of MB-SGD performs well and is widely used, it faces some challenges, such as determining the learning rate without prior knowledge and the risk of converging to suboptimal local minima in non-convex objective functions. To address these, we incorporate the adaptive moment estimation (Adam) algorithm~\cite{Kingma2014} into the MB-SGD. Adam is a momentum-based method featuring adaptive learning rates. The Adam update rule is defined by:
\begin{eqnarray}
	\boldsymbol{m}_{t} &=& \beta_1 \boldsymbol{m}_{t-1} + \left(1-\beta_1\right) \boldsymbol{g}_{t},\label{eq:S20} \\
	\boldsymbol{v}_{t} &=& \beta_2 \boldsymbol{v}_{t-1} + \left(1-\beta_2\right) \left(\boldsymbol{g}_t\odot\boldsymbol{g}_t\right),\label{eq:S21} \\
	\hat{\boldsymbol{m}}_t &=& \frac{\boldsymbol{m}_t}{1-\beta_1^t},\label{eq:S22} \\
	\hat{\boldsymbol{v}}_t &=& \frac{\boldsymbol{v}_t}{1-\beta_2^t},\label{eq:S23} \\
	\boldsymbol{\theta}_{t} &=& \boldsymbol{\theta}_{t-1} - \eta \frac{\hat{\boldsymbol{m}}_t}{\sqrt{\hat{\boldsymbol{v}}_t}+\epsilon}.\label{eq:S24}
\end{eqnarray}
Here, $\boldsymbol{g}_{t}^2$ represents the Hadamard product of $\boldsymbol{g}_{t}$ with itself. $\boldsymbol{m}_{t}$ and $\boldsymbol{v}_{t}$ is the first and second moment vectors, respectively. $\epsilon$ is a small non-zero constant to prevent division by zero. Here, we refer to this optimizer, utilizing MB-SGD with Adam, as MBAdam.

For the two QNN classifiers reported in the main text we used MBAdam with hyperparameters $\beta_1=0.9$ and $\beta_2=0.999$. Learning rates and batch sizes were set according to dataset and task: for EMNIST we employed $\eta=0.1$ and batch size $|B|=100$; for LCEI we used $\eta=0.03$ and $|B|=50$. These configurations were held constant between clean and adversarial training to isolate the effect of adversarial augmentation.

\section{Masked adversarial attack}\label{sec:Mask_FGSM}

\subsection{Numerical simulation of localized attack}

In classical adversarial learning, many algorithms for generating adversarial examples have been proposed. These methods are commonly categorized as white-box attacks or black-box attacks based on the attacker's knowledge of the target model. Among white-box attacks, the fast gradient sign method (FGSM)~\cite{Goodfellow2014} remains a canonical baseline: given a legitimate input $\boldsymbol{x}$, an adversarial example is formed as
\begin{eqnarray}\label{eq:FGSM}
	\boldsymbol{x}' = \boldsymbol{x} + \epsilon \cdot \operatorname{sign}({\nabla}_{\boldsymbol{x}} \mathcal{L}).
\end{eqnarray}
where $\epsilon$ controls the perturbation magnitude and $\nabla_{\boldsymbol{x}} \mathcal{L}$ denotes the input gradient of the loss. Extending such attacks to QNNs, however, is nontrivial: obtaining reliable input gradients on quantum hardware requires repeated circuit evaluations (e.g. via PSR or finite-difference procedures) and thus the cost of generating a single high-dimensional adversarial sample scales unfavourably with the input dimension $N_x$. Concretely, for a shot budget of $N_x$ per circuit evaluation, the naive white-box gradient evaluation scales as $O(2 N_s N_x)$ circuit executions per sample, which becomes prohibitive when attacking high-resolution inputs or when batched generation is desired.

\begin{figure*}[!ht]
	\centering
	\includegraphics[width=0.8\textwidth]{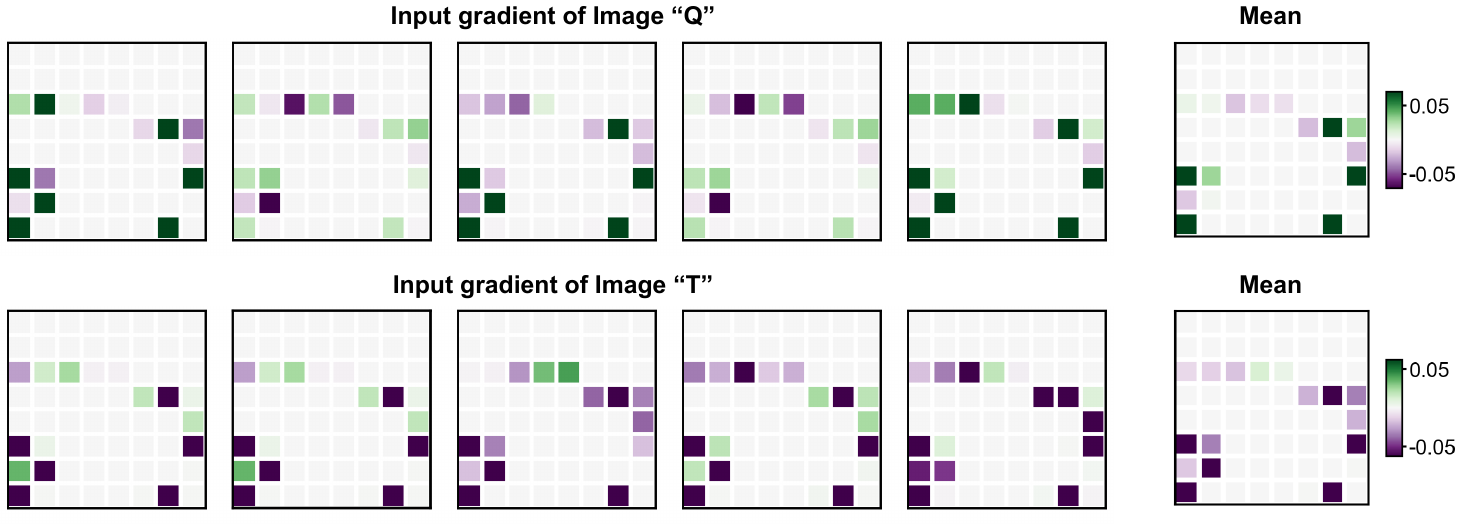}
	\caption{\textbf{Numerical input gradients for EMNIST.} The figures show input gradients computed for five distinct handwritten letter images (``Q'' and ``T''), and the mean input gradient averaged across $20$ different images.
	}
	\label{fig:simu_input_grad}
\end{figure*}

To mitigate this bottleneck, we adopt and analyse a localized (masked) attack strategy that targets a sparse subset of input features selected according to input-gradient magnitudes. We motivate this approach with controlled numerical experiments on a simulated $12$-qubit QNN trained on the same EMNIST task used in the main text. The simulated model mirrors our experimental ansatz (cf. Fig.~\ref{fig:circuit_emnist}) and operates on compressed 
$10 \times 10$ images encoded via an interleaved block scheme. For each test sample, we compute the input gradient vector and examine its statistical structure across the dataset (Fig.~\ref{fig:simu_input_grad}). Two salient observations emerge: (i) the input-gradient vectors are sparse in magnitude and (ii) their large components concentrate in a small, consistently located subset of features across samples. These observations motivate restricting FGSM perturbations to a small, salient subset of features.

\begin{figure*}[!ht]
	\centering
	\includegraphics[width=1\textwidth]{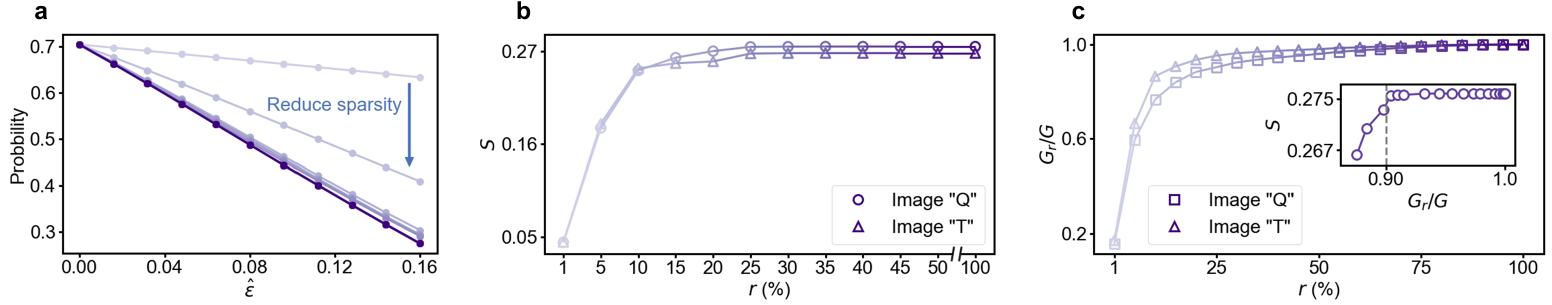}
	\caption{\textbf{Numerical results for local attacks with varying proportion of attacked pixels.} $\textbf{a}$. QNN output as a function of normalized perturbation $\hat{\epsilon}$. $\textbf{b}$. $S_{\Delta p}$ under different attack proportions, computed from the data in panel $\textbf{a}$. $\textbf{c}$. Gradient magnitude proportion $G_r/G$ as a function of pixel proportion. The inset shows $S_{\Delta p}$ as a function of $G_r/G$, where $S_{\Delta p}$ reaches saturation at $90\%$ of the gradient components.
	}
	\label{fig:simu_sparsity}
\end{figure*}

To validate the robustness of the attack to variations in mask size, we performed a systematic ablation study by sweeping the attack sparsity ratio $r$ from 0 to 1. This also facilitates a direct comparison between Mask-FGSM and the non-masked FGSM. Fig.~\ref{fig:simu_sparsity}\textbf{a}, \textbf{b} reports the simulated QNN outputs and corresponding $S_{\Delta p}$ as $r$ varies: $S_{\Delta p}(r)$ rises rapidly and saturates, with approximately the top $25\%$ of pixels already achieving attack efficacy comparable to full-pixel perturbations. 

To quantify how attack sparsity interacts with the empirical distribution of input-gradient magnitudes under realistic noise, we introduce and analyze the cumulative $L_1$-mass statistic $G_r/G$. Let $\overline{\mathcal{G}}=(g_1,g_2,...,g_{N_x})^T$ denote the dataset-averaged input-gradient vector sorted in non-increasing order of absolute value, so that $|g_1| > |g_2| >...> |g_{N_x}| $, The fraction of the total $L_1$-norm captured by the top $r$fraction of pixels as
\begin{eqnarray}\label{eq:G_r}
	\frac{G_r}{G} = \frac{\sum_{i =1}^{r \cdot N_x} {|g_i|}}{||\overline{\mathcal{G}}||_1},
\end{eqnarray}
This quantity measures the degree to which gradient mass concentrates in the most influential input features and thereby provides a principled metric for candidate mask selection. We evaluated $G_r/G$  in numerical experiments that incorporate empirically motivated measurement noise models (shot noise and readout errors matched to our device statistics). Fig.~\ref{fig:simu_sparsity}\textbf{c} plots the resulting $G_r/G$ curve: a pronounced inflection (elbow) is observed near $r \approx 0.25$, beyond which the incremental gain in captured gradient mass decays rapidly. This behaviour reflects the transition from a regime dominated by a compact set of high-signal gradient components to one dominated by many small components whose amplitudes are comparable to measurement noise.

To connect this concentration metric to attack efficacy, we compared $G_r/G$ to the sensitivity $S_{\Delta p}$  computed for corresponding sparsity levels. The inset of Fig.~\ref{fig:simu_sparsity}\textbf{c} shows a tight empirical correlation between $S_{\Delta p}$ and $G_r/G$ across samples and noise realizations: $S_{\Delta p}$  rapidly approaches its asymptote once $G_r/G \approx 0.9$. In operational terms, this implies that selecting the top-$r$ coordinates to satisfy $G_r/G \geq 0.9$ yields attack performance that is statistically indistinguishable from full-input attacks, while reducing gradient-estimation cost by approximately a factor $1/r$. 

Taken together, these results justify a masked strategy for experimental QNN attacks: by restricting perturbations to a small, gradient-informed subset of input features, one obtains nearly saturated adversarial effectiveness at a dramatically reduced measurement and computation cost. Importantly, the masking threshold can be chosen adaptively using $G_r/G$ as an empirical criterion, thereby balancing attack potency against hardware resources and measurement noise. This approach enables practical, batched adversarial generation on NISQ processors and underpins the scalable robustness benchmarking protocol presented in this work.

\subsection{Experimental Mask-FGSM}

\begin{figure*}[!ht]
	\centering
	\includegraphics[width=0.8\textwidth]{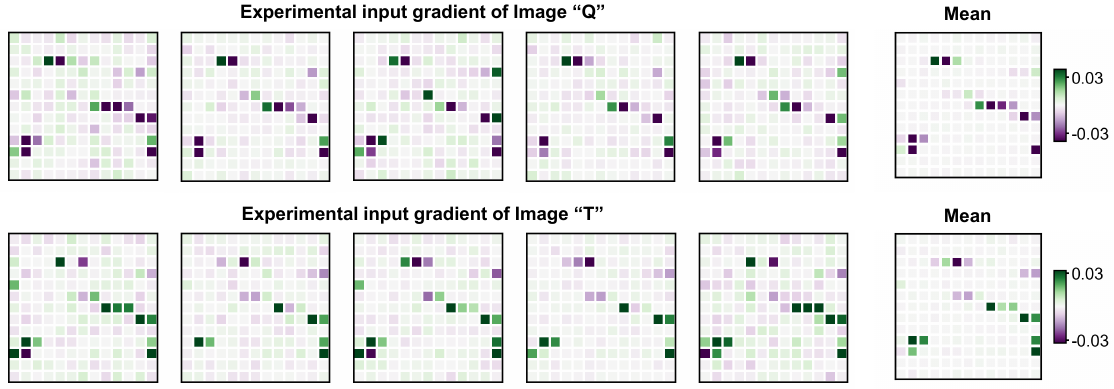}
	\caption{\textbf{Experimental results of the input gradients for EMNIST.} The figures show the experimental input gradients measured for five distinct handwritten letter images (``Q'' and ``T''), and the mean input gradient averaged across $20$ different images.
	}
	\label{fig:exp_input_grad}
\end{figure*}

We experimentally evaluated input gradients on our superconducting processor for a representative set of samples. The results are displayed in Fig.~\ref{fig:exp_input_grad}. The measured gradient fields exhibit consistent salient features across distinct images, with a small subset of input coordinates accounting for a large fraction of the total gradient mass. Motivated by this empirical sparsity, we introduce the masked attack (Mask-FGSM): a localized white-box attack that restricts perturbations to a preselected subset of input features determined by a binary mask $\mathcal{M} \in \{0,1\}^{\operatorname{dim}(\boldsymbol{x})}$. The Mask-FGSM is designed to (i) minimize the number of perturbed pixels (thereby reducing perceptual footprint), (ii) substantially cut the number of input-gradient evaluations required to construct adversarial examples, and (iii) mitigate the deleterious influence of measurement noise by excluding low-magnitude, noise-dominated gradient components from the attack support.

\begin{figure*}[!ht]
	\centering
	\includegraphics[width=0.6\textwidth]{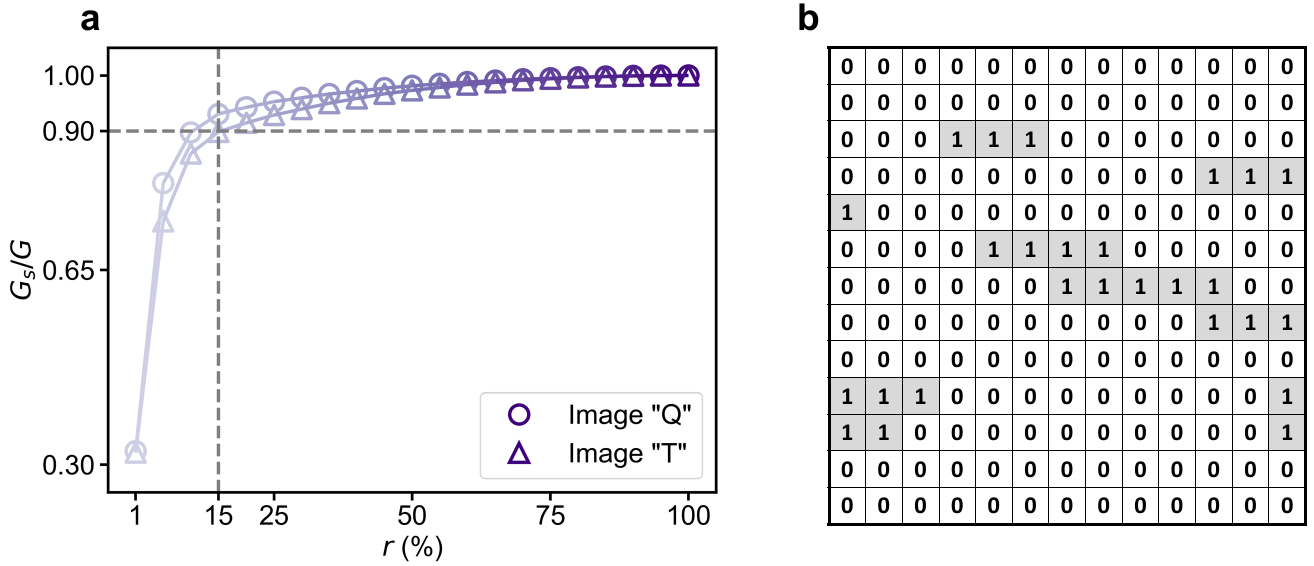}
		\caption{\textbf{Mask $\mathcal{M}$ construction based on experimental gradient sparsity for EMNIST.} $\textbf{a}$, Gradient magnitude proportion $G_r/G$ as a function of pixel proportion, illustrating that the top $15\%$ of pixels yield approximately $90\%$ of the total gradient magnitude. $\textbf{b}$, The binary mask $\mathcal{M}$ used in EMNIST experiments, where the elements corresponding to the top $15\%$ of pixels are assigned $1$, and all other positions are assigned $0$.
	}
	\label{fig:exp_sparsity}
\end{figure*}

The mask $\mathcal{M}$ is constructed via a data-driven procedure. Experimentally, we estimate $\overline{\mathcal{G}}$ by averaging single-sample gradients over a calibration set of $20$ images. Fig.~\ref{fig:exp_sparsity}\textbf{a} reports $G_r/G$ as a function of $r$; the curve exhibits a pronounced elbow near $r \approx 0.15$, at which point $G_r/G \approx 0.9$. Accordingly, for EMNIST, we set $r = 0.15$  and define  $\mathcal{M}$ by activating indices corresponding to the top $15\%$ gradient components (entries set to 1) and zeroing the remainder. The spatial structure of the resulting mask is illustrated in Fig.~\ref{fig:exp_sparsity}\textbf{b} and was used across attack runs on EMNIST (see Algorithm~\ref{Algo_emnist}).


\begin{figure*}[!ht]
	\centering
	\includegraphics[width=1\textwidth]{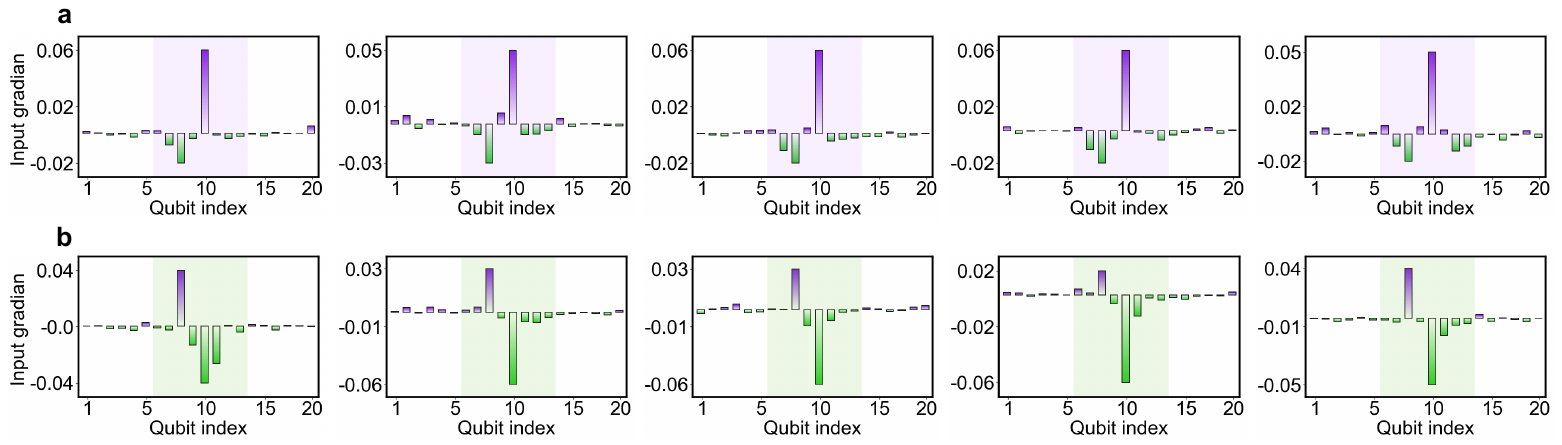}
	\caption{\textbf{Experimental input gradients for LCEI.} The figures show the experimental input gradients for five distinct non-excited ($\textbf{a}$) and excited cluster states ($\textbf{b}$). Significant gradient components corresponding to rotation angles are observed on the central qubits ($\rm{Q7}-\rm{Q14}$), which are highlighted with purple and green backgrounds. Each panel shows the average results obtained from 5 experimental repetitions.
	}
	\label{fig:lcei_input_grad}
\end{figure*}

An analogous procedure was adopted for LCEI. We computed input gradients with respect to the rotation-angle parameter $\alpha$ of each of the $20$ $R_x$ gates that generate cluster-state excitations (Fig.~\ref{fig:lcei_input_grad}). The measured gradients show a pronounced concentration on a contiguous subset of qubits, notably qubits $\rm{Q7}$ to $\rm{Q14}$, across different cluster-state samples. Guided by this observation, we constructed a rotation-angle mask that restricts perturbations to the $\alpha$ angles of $\rm{Q7}-\rm{Q14}$ and executed localized attacks in this reduced parameter subspace (Algorithm~\ref{Algo_LCEI}). This targeted strategy reproduces the computational savings and noise-robustness benefits observed for EMNIST.

\begin{algorithm}[H]
	\caption{Mask FSGM of QNN classifier for EMNIST dataset}
	\label{Algo_emnist}
	\begin{algorithmic}[1]
		\REQUIRE A well-trained QNN classifier with parameters $\boldsymbol{\theta}^{*}$, the loss function $\mathcal{L}$, the step size of the perturbation $\epsilon$, the binary mask vector $\mathcal{M}=(m_1,m_2,...,m_n)^T$, and a legitimate image $\boldsymbol{x}=(x_1,x_2,...,x_n)^T$ with label $\boldsymbol{c}$.
		\ENSURE The adversarial image $\boldsymbol{x}'$
		\STATE Initialization: $\boldsymbol{x}' \leftarrow \boldsymbol{x}$, $\boldsymbol{\delta} \leftarrow \boldsymbol{0} $
		\FOR{ $i \in [n]$}
		\IF {$m_i=1$}
		\STATE Calculate the gradients of loss $\mathcal{L}$ with respect to component $x_i$: $\frac{\partial \mathcal{L}(f(\boldsymbol{x};\boldsymbol{\theta}^{*}),\boldsymbol{c}) }{\partial x_i }$
		\STATE Generate perturbation: ${\delta}_{i} \leftarrow \epsilon \cdot \operatorname{sgn}\left[ \frac{\partial \mathcal{L}(f(\boldsymbol{x};\boldsymbol{\theta}^{*}),\boldsymbol{c}) }{\partial x_i } \right]$
		\STATE Updates: ${x}_i' \leftarrow {x}_i + {\delta}_{i}$
		\ENDIF
		\ENDFOR
		\RETURN $\boldsymbol{x}'$
	\end{algorithmic}
\end{algorithm}

\begin{algorithm}[H]
	\caption{Mask FSGM of QNN classifier for LCEI}
	\label{Algo_LCEI}
	\begin{algorithmic}[1]
		\REQUIRE A well-trained QNN classifier with parameters $\boldsymbol{\theta}^{*}$, the loss function $\mathcal{L}$, the step size of the perturbation $\epsilon$, the binary mask vector $\mathcal{M}=(m_1,m_2,...,m_n)^T$, a legitimate $n$-qubits linear cluster state ${\rho}_{\rm{in}}$ with the rotation angle vector $\boldsymbol{\alpha}$ of $R_x$ gates, and label $\boldsymbol{c}$.
		\ENSURE The adversarial quantum state ${\rho}'_{\rm{in}}$ and corresponding rotation angle vector $\boldsymbol{\alpha}'$
		\STATE Initialization: $\boldsymbol{\alpha}' \leftarrow  \boldsymbol{\alpha}$, $\boldsymbol{\delta} \leftarrow \boldsymbol{0} $
		\FOR{ $i \in [n]$}
		\IF {$m_i=1$}
		\STATE Calculate the gradients of loss $\mathcal{L}$ with respect to angle ${\alpha}_i$: $\frac{\partial \mathcal{L}(f(\rho_{\rm{in}};\boldsymbol{\theta}^{*}),\boldsymbol{c}) }{\partial {\alpha}_i }$
		\STATE Generate perturbation: ${\delta}_{i} \leftarrow \epsilon \cdot \operatorname{sgn}\left[\frac{\partial \mathcal{L}(y(\rho_{\rm{in}};\boldsymbol{\theta}^{*}),\boldsymbol{c}) }{\partial {\alpha}_i } \right]$
		\STATE Updates: ${\alpha}_i' \leftarrow {\alpha}_i + {\delta}_{i}$
		\ENDIF
		\ENDFOR
		\RETURN $\rho'_{\rm{in}}$ and $\boldsymbol{\alpha}'$
	\end{algorithmic}
\end{algorithm}

\section{Robustness bound of QNN classifier}

\subsection{Robustness lower bound}

In the main text, we present a method for extracting the robustness upper bound through adversarial attacks. Here, we provide the proof for the robustness lower bound given in Eq.~(\ref{eq: R_LB}) in the main text. For a given input quantum state $\rho_{\mathrm{in}}$, the QNN together with the final measurement on the output qubit 
defines a completely positive trace-preserving (CPTP) map $\mathcal{E}$ from the full input Hilbert space to the reduced
state of the output qubit. We denote by
\begin{equation}
\rho := \rho_{\mathrm{out}} = \mathcal{E}(\rho_{\mathrm{in}}), \qquad
\sigma := \sigma_{\mathrm{out}} = \mathcal{E}(\sigma_{\mathrm{in}}),
\end{equation}
the reduced density matrices of the output qubit corresponding to the legitimate and perturbed input states, respectively.
All quantities below are defined on this output Hilbert space.

We consider the probability distributions induced by measuring the output qubit with a POVM 
$\{\Pi_k\}$ corresponding to the classifier readout, where
\begin{equation}
p_k = \operatorname{tr}(\Pi_k \rho), \qquad 
q_k = \operatorname{tr}(\Pi_k \sigma).
\end{equation}
We define the vectors $\boldsymbol{p}$ and $\boldsymbol{q}$ with components $\sqrt{p_k}$ and $\sqrt{q_k}$, respectively.
Their inner product $\boldsymbol{p} \cdot \boldsymbol{q} = \sum_k \sqrt{p_k q_k}$ is related to the quantum fidelity $F(\rho,\sigma)$. According to quantum measurement theory, this relationship is governed
by the inequality
\begin{eqnarray}\label{eq:fidelity_inequality}
\boldsymbol{p} \cdot \boldsymbol{q} \ge \sqrt{F(\rho,\sigma)}.
\end{eqnarray}

To derive the robustness lower bound, we seek the maximum possible value of $\boldsymbol{p} \cdot \boldsymbol{q}$ 
under the condition that the state $\sigma$ is just classified incorrectly. 
Assuming the elements of $\boldsymbol{p}$ are sorted in descending order ($p_1 > p_2 > \cdots > p_n$), this problem can be formulated as
\begin{eqnarray}
\begin{aligned}
\operatorname{max} \quad & \boldsymbol{p} \cdot \boldsymbol{q}, \\
\text{s.t.} \quad & \|\boldsymbol{q}\|_2 = 1, \\
& \prod_{i=2}^n (q_1 - q_i) = 0,
\end{aligned}
\end{eqnarray}
where the constraint $\prod_{i=2}^n (q_1 - q_i) = 0$ enforces the misclassification threshold, i.e., the probability
associated with the true class is no longer strictly larger than all competing classes. Solving this constrained optimization using the Lagrange multiplier method (see Ref.~\cite{Guan2020}) yields the optimal
value at the threshold $q_1 = q_2$,
\begin{equation}
V^* = \sqrt{1 - \frac{1}{2}(\sqrt{p_1} - \sqrt{p_2})^2},
\end{equation}
where $p_1$ and $p_2$ denote the largest and second-largest components of $\boldsymbol{p}$, respectively. Therefore, if
\begin{equation}
D(\rho,\sigma) = 1 - F(\rho,\sigma) < \frac{1}{2}(\sqrt{p_1} - \sqrt{p_2})^2,
\end{equation}
then $\sqrt{F(\rho,\sigma)} > V^*$. By Eq.~(\ref{eq:fidelity_inequality}), this implies
$\boldsymbol{p} \cdot \boldsymbol{q} > V^*$, which guarantees that $\boldsymbol{q}$ lies outside the misclassified region.
Therefore, the perturbed state $\sigma$ is always correctly classified, and the robustness lower bound is given by
\begin{equation}
R_{\rm LB} = \frac{1}{2}(\sqrt{p_1} - \sqrt{p_2})^2.
\end{equation}

\subsection{Validity of the bounds under experimental noise}

The analytical lower bound $R_{\rm{LB}}$ utilized in this work is derived from the relationship between the Bures distance and the statistical distance of probability distributions. A key question arises regarding its applicability to experimental implementations where states are mixed due to decoherence and measurements are distorted by readout noise.

The derivation of Eq.~(\ref{eq: R_LB}) in the main text rests on the inequality provided in Eq.~(\ref{eq:fidelity_inequality}) of Section IV.A. It is a well-established result in quantum information theory (Fuchs-Caves inequalities ~\cite{PhysRevLett.73.3047}) that this relation holds for arbitrary density matrices $\rho$ and $\sigma$, and for any POVM.


In our experiments, the classifier is defined by the physical hardware operation. The hardware noise, such as amplitude damping and depolarization, transforms the ideal pure state into a mixed state $\rho_{\rm exp}$. Similarly, readout noise modifies the ideal projective measurement $\{\Pi_k\}$ into an effective noisy POVM $\{E_k\}$. The experimentally measured probabilities $\tilde{p}_k = \operatorname{Tr}(\rho_{\rm exp} E_k)$ incorporate these effects. Therefore, the non-unitary operations inherent in the NISQ circuit do not invalidate the mathematical inequality; rather, the bound certifies the robustness of the actual physical state prepared on the processor, accounting for system noise.

\subsection{Impact of error mitigation on the bounds}

In experimental implementations, measurement results are corrupted by readout errors, which can be modeled by a confusion matrix acting on the ideal probability vector. To benchmark the intrinsic robustness of the QNN circuit, which arises from the variational ansatz and quantum noise like depolarization, rather than the performance of the readout signal chain, we apply readout error mitigation (REM) using iterative bayesian unfolding (IBU), as detailed in Section VII.C.
This necessitates a careful interpretation of the robustness bounds:

\begin{itemize}
    \item Lower bound $R_{\rm LB}$: Experimental readout noise typically blurs the probability distribution, artificially reducing the prediction margin $|p_1 - p_2|$. By applying REM, we recover the probabilities corresponding to the actual quantum state prepared by the circuit, stripping away the classical confusion introduced by the measurement chain. Mathematically, this sharpens the distribution and increases the calculated $R_{\rm LB}$. This is desirable, as our goal is to benchmark the robustness of the QNN itself, rather than the readout electronics.

    \item Upper bound $R_{\rm UB}$: The upper bound is derived from the state infidelity $D(\rho,\sigma)$. The density matrices $\rho$ and $\sigma$ were reconstructed via quantum state tomography (QST). Our QST reconstruction procedure via MLE incorporates the calibration of measurement operators (readout fidelities). Therefore, the reconstructed state $\rho$ represents the physical quantum state before readout, effectively decoupling readout errors from the state's intrinsic decoherence noise.
\end{itemize}

By applying EM, implicitly in QST for $R_{\rm UB}$ and explicitly via IBU for $R_{\rm LB}$, we align both metrics to the same physical reference: the noisy mixed state generated by the circuit measured by an ideal projective measurement. The validity of the bound holds for this reference frame: $p_k$ represents the mitigated probability of the noisy state $\rho$. The remarkably small gap ($\sim 10^{-3}$) between the analytical $R_{\rm LB}$ and empirical $R_{\rm UB}$ observed in our data serves as strong empirical evidence that this consistent treatment of readout errors effectively aligns the circuit-level robustness.

\section{Noise-Induced Sensitivity Scaling}
In this section, we focus on how decoherence noise enhances the robustness of QNN classifiers. Let the output expectation value of the Pauli-$z$ operator in a QNN be denoted by
\begin{eqnarray}\label{eq:G_r}
    z(\boldsymbol{x}):=\langle \sigma_z \rangle (\boldsymbol{x}),
\end{eqnarray}
for an input sample $\boldsymbol{x}$. For a perturbed input $\boldsymbol{x}+\boldsymbol{\delta}$ with normalized perturbation strength $\hat{\epsilon}$, the noiseless baseline deviation of the observable is
\begin{eqnarray}\label{eq:G_r}
    \Delta z(\boldsymbol{x}) = z(\boldsymbol{x}+\boldsymbol{\delta}) - z(\boldsymbol{x})
\end{eqnarray}
for the quantum logit defined in the main text
\begin{eqnarray}\label{eq:G_r}
    L(\boldsymbol{x}) = \operatorname{ln} \left( \frac{1 + z(\boldsymbol{x}) }{1 - z(\boldsymbol{x}) } \right). 
\end{eqnarray} 
For small perturbations, a first-order Taylor expansion around $z(\boldsymbol{x})$ yields
\begin{eqnarray}\label{eq:G_r}
    \Delta L = L(z+\Delta z) - L(z) \approx \frac{dL}{dz}  \Delta z + O((\Delta z)^2),
\end{eqnarray} 
with ${dL}/{dz}={2}/(1-z^2)$. Thus, to first order,
\begin{eqnarray}\label{eq:G_r}
    \Delta L(\boldsymbol{x}) \approx \frac{2 \Delta z(\boldsymbol{x})}{1-z(\boldsymbol{x})^2}.
\end{eqnarray} 

Any single-qubit state can be written in the Bloch representation as $\rho = \frac{1}{2}(I+\boldsymbol{r} \cdot \boldsymbol{\sigma} )$, with Bloch vector $\boldsymbol{r}$. A general single-qubit CPTP channel $\Lambda$ has a Pauli transfer matrix (PTM) representation
\begin{eqnarray}\label{eq:G_r}
    R = \begin{pmatrix}
		1 &  \boldsymbol{0}\\
		\boldsymbol{\chi} & \tilde{R}
	\end{pmatrix}, \quad \tilde{R} \in \mathbb{R}^{3 \times3}
\end{eqnarray}
where $\tilde{R} = \operatorname{diag}(R_{xx}, R_{yy}, R_{zz})$ acts on local Bloch vector, and $\boldsymbol{\chi}$ encodes non-unital shift. For amplitude damping ($T_1$) and pure dephasing ($T_2$) channels over a time interval $t$, 
\begin{eqnarray}\label{eq:G_r}
    \tilde{R}_{\rm{AD}} = \operatorname{diag}\left(\sqrt{1- \gamma_1},\sqrt{1-\gamma_1},1-\gamma_1\right), \quad \boldsymbol{\chi}_{\rm{AD}}=(0,0,\gamma_1)^T,
\end{eqnarray}
\begin{eqnarray}\label{eq:G_r}
    \tilde{R}_{\rm{PD}} = \operatorname{diag}\left(1-\gamma_2, 1-\gamma_2, 1\right), \quad \boldsymbol{\chi}_{\rm{PD}}=\boldsymbol{0},
\end{eqnarray}
where $\gamma_1=1-e^{-t/T_1}$ and $\gamma_2=1-e^{-t/T_2}$. Their composition yields
\begin{eqnarray}\label{eq:R_comp}
    \tilde{R} = \operatorname{diag}\left((1-\gamma_2) \sqrt{1- \gamma_1},(1-\gamma_2)\sqrt{1-\gamma_1},1-\gamma_1\right), \quad \boldsymbol{\chi} =(0,0,\gamma_1^T
    ).
\end{eqnarray}

Consider an $n$-qubit QNN circuit composed of $l$ alternating layers of single-qubit $SU(2)$ gates and two-qubit CZ gates (like Fig.~\ref{fig:circuit_emnist} and Fig.~\ref{fig:LCEI}). Such circuits approximate a unitary $2$-design, implying that from the perspective of any local qubit, the effective action of noise is statistically equivalent to Haar-random conjugation by $U \in U(2)$. Define the twirled channel of $\Lambda$ as
\begin{eqnarray}\label{eq:G_r}
    \overline{\Lambda} = \int_{U \in U(2)} U^{\dagger} \Lambda (U \rho U^{\dagger} ) U d \mu (U) ),
\end{eqnarray}
where $ d\mu$ is the Haar measure.

By Schur’s lemma, any linear map invariant under conjugation by the full unitary group must be a linear combination of the identity channel and the completely depolarizing channel~\cite{NIELSEN2002249}. Hence
\begin{eqnarray}\label{eq:G_r}
    \overline{\Lambda} = (1-\xi)\rho + \xi \frac{I}{2},
\end{eqnarray}
with a depolarizing factor
\begin{eqnarray}\label{eq:xi}
    1-\xi = \frac{\operatorname{tr}(\tilde{R})}{3} = \frac{R_{xx}+ R_{yy}+ R_{zz}}{3}.
\end{eqnarray}
The non-unital shift $\boldsymbol{\chi}$ averages to zero under the Haar measure.

Substitute (\ref{eq:R_comp}) into (\ref{eq:xi}),
\begin{eqnarray}\label{eq:xi}
    \xi = 1-\frac{2\left((1-\gamma_2) \sqrt{1-\gamma_1}\right)+(1-\gamma_1)}{3} = 1-\frac{2e^{-t \left( \frac{1}{2T_1} + \frac{1}{T_2}\right)} + e^{-t/T_1}}{3}.
\end{eqnarray}
For $t \ll T_1,T_2$, expanding the exponentials,
\begin{eqnarray}\label{eq:xi}
    \xi \approx \frac{2}{3}\left( \frac{1}{T_1} + \frac{1}{T_2} \right)t + O(t^2).
\end{eqnarray}
After $l$ layers, the Bloch vector contracts as $\boldsymbol{r}_l \approx (1-\xi)^l \boldsymbol{r}_0 \Rightarrow z_l \approx (1-\xi)^l z_0$. Perturbations scale as
\begin{eqnarray}\label{eq:xi}
    \Delta z_l \approx (1-\xi)^l \Delta z_0.
\end{eqnarray}
Hence, the noisy sensitivity is
\begin{eqnarray}\label{eq:xi}
    S^{\rm{noisy}}=\frac{2 \Delta z_l}{\hat{\epsilon}\left(1-z_l^2\right)} = \frac{2 (1-\xi)^l \Delta z_0}{\hat{\epsilon}\left(1- (1-\xi)^{2l}  z_0^2\right)} = \underbrace{\frac{ (1-\xi)^l \left( 1-z_0^2 \right)}{1-(1-\xi)^{2l}   z_0^2}}_{:=C(\xi,l,z_0)} S^{\rm{ideal}}.
\end{eqnarray}
where $S^{\rm{ideal}}$ is the sensitivity in the noiseless condition. The above equation provides the exact first-order ratio factor $C(\xi,l,z_0)$, which scales the noiseless sensitivity to the noisy situation.

Let $u= (1-\xi)^l$, $u \in [0,1]$. We now prove that $S^{\rm{noisy}} \leq S^{\rm{ideal}}$ is equivalent to proving
\begin{eqnarray}\label{eq:xi}
    C(\xi,l,z_0)=\frac{u\left( 1-z_0^2 \right)}{1-u^{2}z_0^2} \leq 1.
\end{eqnarray}
Define
\begin{eqnarray}\label{eq:xi}
    f(u):=u^2 z_0^2+ u \left(1-z_0^2\right) - 1.
\end{eqnarray}
Note $f(1)=0$, and
\begin{eqnarray}\label{eq:xi}
    f'(u)=2z_0^2u + \left(1-z_0^2\right) \geq 1-z_0^2 \geq 0.
\end{eqnarray}
Thus, $f(u)$ is monotonically non-decreasing on 
$[0,1]$. Since $f(1)=0$, for all $0 \leq u \leq 1$ we have $f(u) \leq 0$, implying $C(\xi,l,z_0) \leq 1$. Therefore
\begin{eqnarray}\label{eq:xi}
S^{\rm{noisy}}<S^{\rm{ideal}},
\end{eqnarray}
with strict inequality whenever $\xi > 0$.
Finally, the robustness score
\begin{eqnarray}\label{eq:xi}
R(\boldsymbol{x})=\frac{1}{1+e^{S(\boldsymbol{x})}},
\end{eqnarray}
is strictly increased under noise. Averaging over the dataset, the adversarial robustness measure $\overline{R}_{\rm{adv}}$ is improved.

It is essential to clarify the physical nature of the robustness observed in our analysis. As pointed out by the derivation above, the primary effect of the noise channel is to scale down the sensitivity $S$ by a factor $C(\xi, l, z_0) < 1$. In the context of adversarial learning, this phenomenon is often referred to as gradient attenuation or gradient masking. Fundamentally, the noisy quantum circuit becomes less responsive to all input changes, not just adversarial ones. This creates a trade-off between robustness and discriminability. The robustness advantage reported in this work exists because the experiment operates in a regime where the noise level is: 1. High enough to attenuate the gradients of microscopic adversarial perturbations, effectively masking the precise direction needed to flip the label; 2. Low enough to preserve the macroscopic signal of the legitimate data features.

This is evidenced by the fact that our QNN retains high classification accuracy despite the noise. If the noise were increased further, the gradient attenuation would become destructive, causing the model to lose its ability to classify clean data. Therefore, the observed robustness should be interpreted as a hardware-induced regularization effect, a ``low-pass filter'' that suppresses high-frequency adversarial noise while passing low-frequency semantic features, rather than an unconditional algorithmic superiority.

\section{Quantum processor}

\subsection{Device information}

Our experiment was executed on a $72$-qubit superconducting processor, featuring a $6 \times 12$ arrangement of transmon qubits in a 2D lattice, divided into $12$ frequency-multiplexed readout groups. Qubits are dispersively coupled to their dedicated readout resonator, which then shares a common readout input line in each frequency-multiplexed readout group. Each qubit exhibits a nonlinearity of around $-240$ MHz, with individual microwave and flux lines for driving excitations and tuning its resonance frequency between $|0 \rangle$ and $|1 \rangle$. We designed two types of qubits with different tunable frequency ranges: high-frequency qubits with a maximum frequency of $4.8$ GHz, and low-frequency qubits being $4.4$ GHz. The processor includes $126$ tunable couplers, enabling adjustable effective coupling strength between neighboring qubits with a maximum of $-40$ MHz. Each coupler is also a transmon qubit, controlled by a single flux line for frequency tuning across a range of approximately $4$ to $6.5$ GHz. Due to fabrication defects, including Josephson junction shorts and internal wire breaks within the packaging box, five qubits and four couplers on our processor are non-functional. 

\begin{figure*}[!ht]
	\centering
	\includegraphics[width=0.6\textwidth]{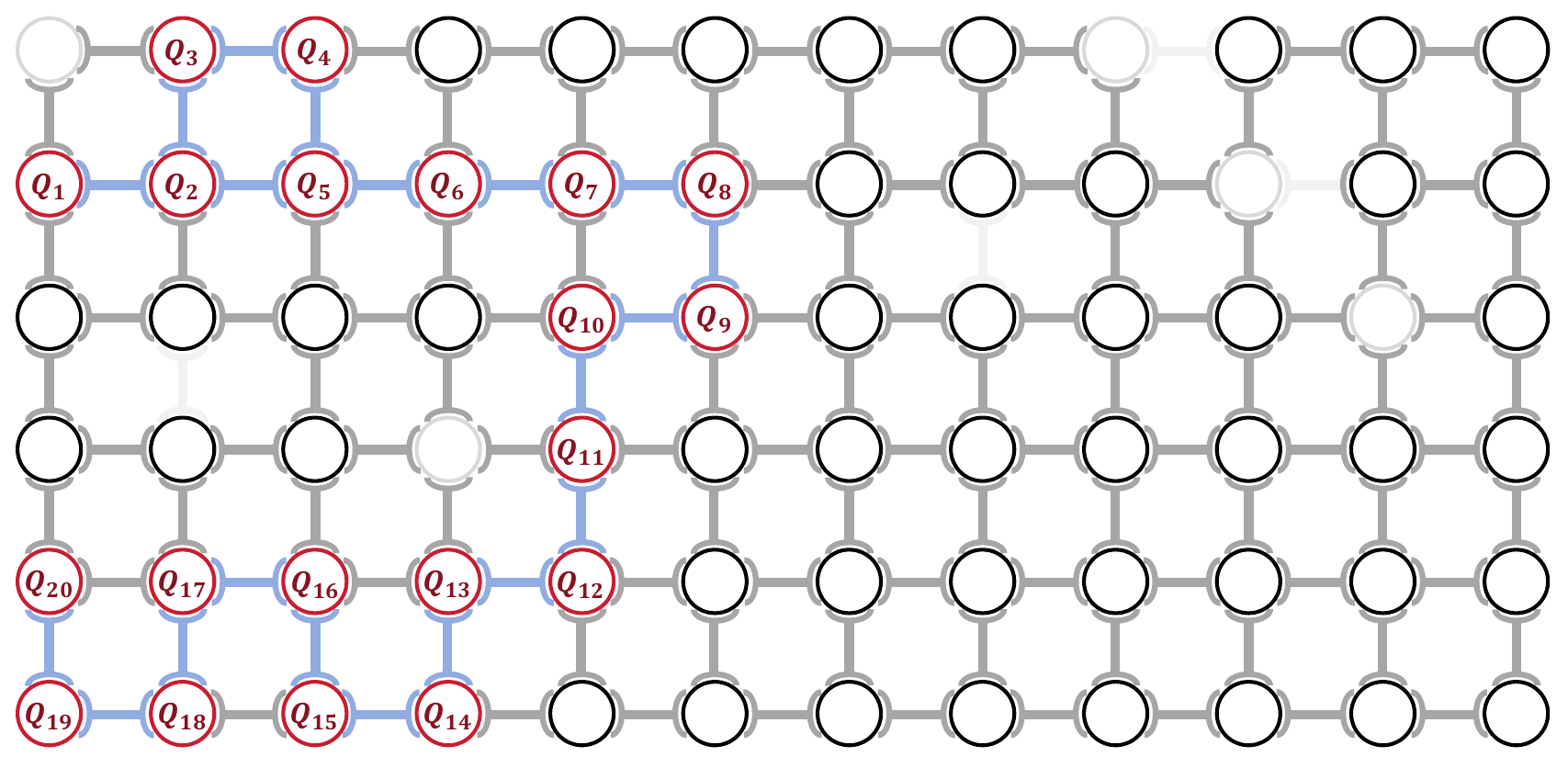}
	\caption{\textbf{Layout of our 72-qubit superconducting processor.} The $20$ operational qubits and their associated couplers utilized in our experiment are highlighted in red and blue, respectively. Non-functional qubits and couplers are indicated by a light gray color.
	}
	\label{fig:chip_topology}
\end{figure*}

\begin{figure*}[!t]
	\centering
	\includegraphics[width=0.9\textwidth]{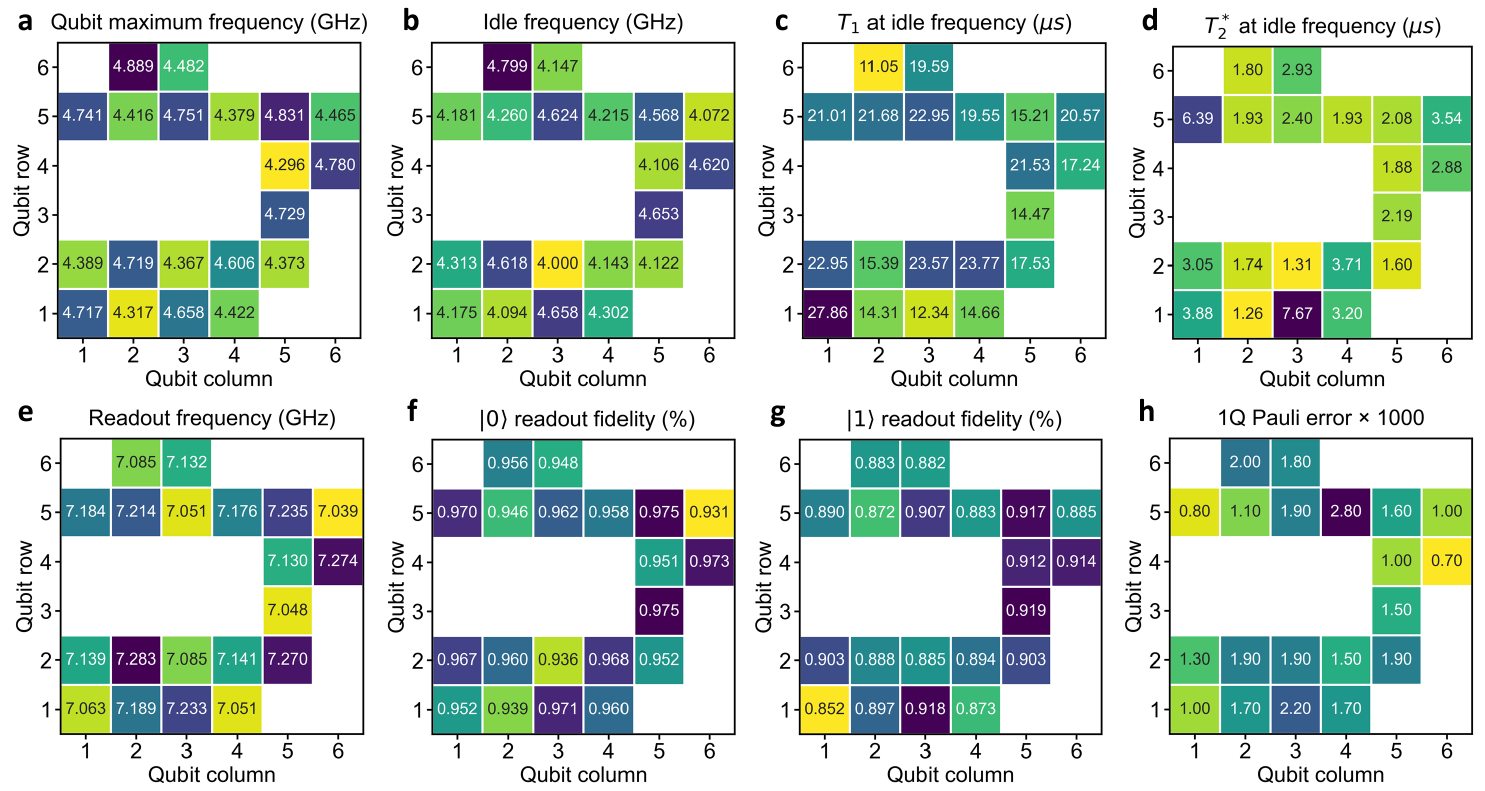}
	\caption{\textbf{Heatmaps of various single-qubit parameters for the $20$ qubits.} $\textbf{a}$, $\textbf{b}$, Qubit maximum frequency and idle frequencies, respectively. $\textbf{c}$, $\textbf{d}$, Qubit energy relaxation and Ramsey dephasing time, respectively, measured at their idle frequencies. $\textbf{e}$, Readout resonator frequencies. $\textbf{f}$, $\textbf{g}$, Simultaneous readout fidelity when qubits are prepared at $|0\rangle $ and $|1\rangle $ states, respectively. $\textbf{h}$, Single-qubit Pauli error characterized by simultaneous XEB experiment.
	}
	\label{fig:1Q_parameters}		
\end{figure*}

For the experiment, we selected $20$ qubits, denoted as $Q_i$, where $i=1,2,\ldots,20$, showing in Fig.~\ref{fig:chip_topology}. Their maximum frequencies and the readout resonator frequencies are shown in Fig.~\ref{fig:1Q_parameters}\textbf{a} and \textbf{e}. The energy relaxation ($T_1$) times and the Ramsey dephasing ($T_2^*$) times measured at the idle frequency are shown in Fig.~\ref{fig:1Q_parameters}\textbf{c} and \textbf{d}. Average characteristic parameters of these $20$ qubits are listed in Table~\ref{tab:tableS1}.

Our superconducting quantum processor is fabricated using flip-chip technology, with all qubits and couplers located on the top chip, control/readout lines and readout resonators situated on the bottom chip. The two chip substrates are physically and electrically interconnected via indium bumps: by precisely aligning the two chips and performing thermocompression bonding, the chips are firmly joined together via the bumps, resulting in a completed flip-chip device. Additionally, a niobium airbridge process is implemented on the chip to suppress parasitic slotline modes in the coplanar waveguides, thereby reducing electromagnetic crosstalk between different control lines. Following chip fabrication, the processor is packaged within a silver-plated aluminum sample box for electromagnetic shielding. Qubit control lines are wire-bonded to one end of PCB strips, with the other end connected to external SMP RF coaxial connectors.

\begin{figure*}[!t]
	\centering
	\includegraphics[width=0.5\textwidth]{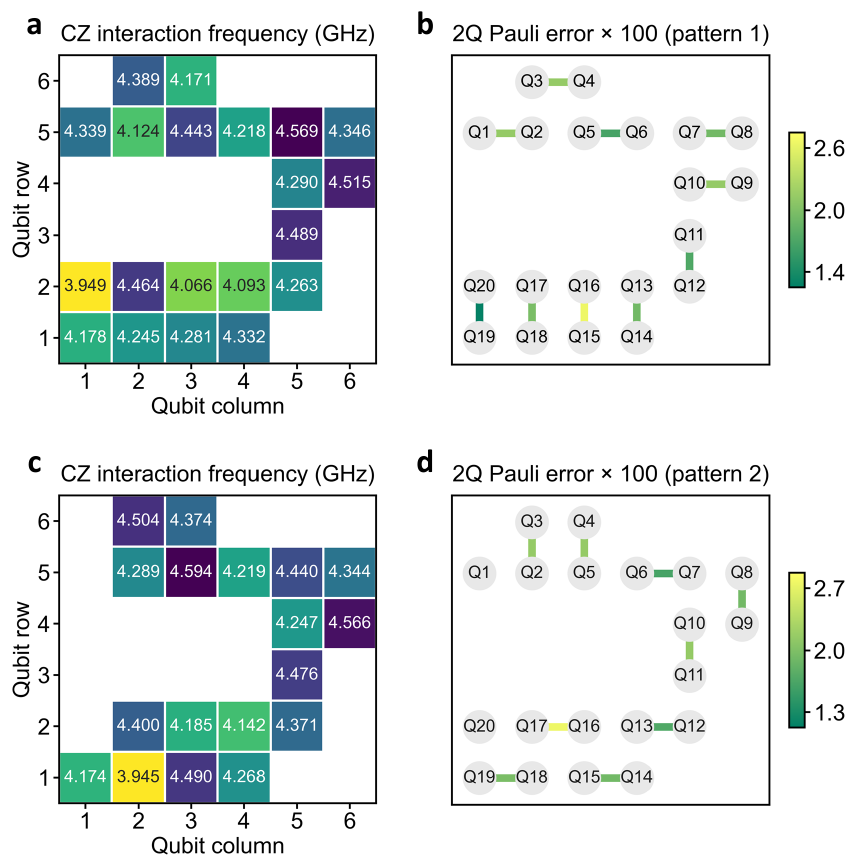}
	\caption{\textbf{Heatmaps of CZ gate parameters.} $\textbf{a}$, $\textbf{c}$, Simultaneous CZ gate interaction frequencies for the two respective patterns. $\textbf{b}$, $\textbf{d}$, Two-qubit Pauli error characterized by simultaneous $2$Q-XEB experiment for the two respective patterns.}
	\label{fig:2Q_parameters}
\end{figure*}

\begin{table*}[!ht]
	\caption{\label{tab:tableS1}%
		System parameters for our quantum processor.
	}
	\begin{ruledtabular}
		\centering
			\begin{tabular}{ccccc}
					Parameters & Median & Mean & Stdev. & Units \\
					\hline
					Qubit maximum frequency & 4.544 & 4.566 & 0.187 & GHz \\
					
					Qubit idle frequency & 4.237 & 4.333 & 0.244 & GHz \\
					
					Qubit anharmonicity & -0.237 & -0.232 & 0.017 & GHz \\
					
					\hline
					
					Readout resonator frequency & 7.140 & 7.151 & 0.080 & GHz\\
					
					Readout linewidth $\kappa/2\pi$ & 1.07 & 1.17 & 0.43 & MHz\\
					
					Dispersive shift $\chi/2\pi$  & 1.15 & 1.14 & 0.21 & MHz\\
					
					Readout fidelity of $|0\rangle$ & 0.959 & 0.958 & 0.013 & \% \\
					
					Readout fidelity of $|1\rangle$ & 0.891 & 0.893 & 0.017 & \% \\
					
					\hline
					
					$T_1$ at idle frequency  & 19.57 & 18.86 & 4.33 & $\mu$s \\
					
					$T_2$ Ramsey at idle frequency & 2.29 & 2.86 & 1.59 & $\mu$s \\
					
					Simultaneous 1Q Pauli error & 0.17 & 0.16 & 0.05 & \% \\
					
					Individual 2Q Pauli error & 1.92 & 1.91 & 0.42 & \%\\
					
					Simultaneous 2Q Pauli error (pattern 1) & 1.98 & 1.99 & 0.40 & \%\\
					
					Simultaneous 2Q Pauli error (pattern 2) & 1.98 & 1.90 & 0.52 & \%\\
					
			\end{tabular}
		\end{ruledtabular}
	\end{table*}

\subsection{Experimental setup}

The quantum processor is installed in a dilution refrigerator (DR) and cooled to a base temperature of $15$ mK. Each qubit has two control lines: the XY line for qubit driving and the Z control line for biasing the qubit to its idle point. Each coupler has one Z control line. Waveform generation is based on custom arbitrary waveform generator (AWG) modules. Each AWG module provides $8$ DACs with a $14$-bit resolution, providing sampling rates of $3.2$ GS/s for XY-pulses and $1.2$ GS/s for Z-pulses. The bias signals and fast Z-pulse for qubits and couplers are directly output by the AWG. Single-qubit XY drive and measurement microwave pulses are synthesized by mixing the AWG output with a local oscillator (LO) signal, producing signals with arbitrary spectral content within a $\pm 250$ MHz bandwidth. For qubit state readout, the input measurement microwave is delivered to the quantum processor through the readout line, and the output signal passes through three circulators before being amplified by an impedance-matched parametric amplifier (IMPA). Then, it passes through a fourth circulator and undergoes further amplification at $4$K and room temperature by high electron mobility transistor (HEMT) amplifiers. Finally, the amplified signal is sampled and processed by room-temperature electronics. 

\begin{figure*}[!th]
	\centering
	\includegraphics[width=0.4\textwidth]{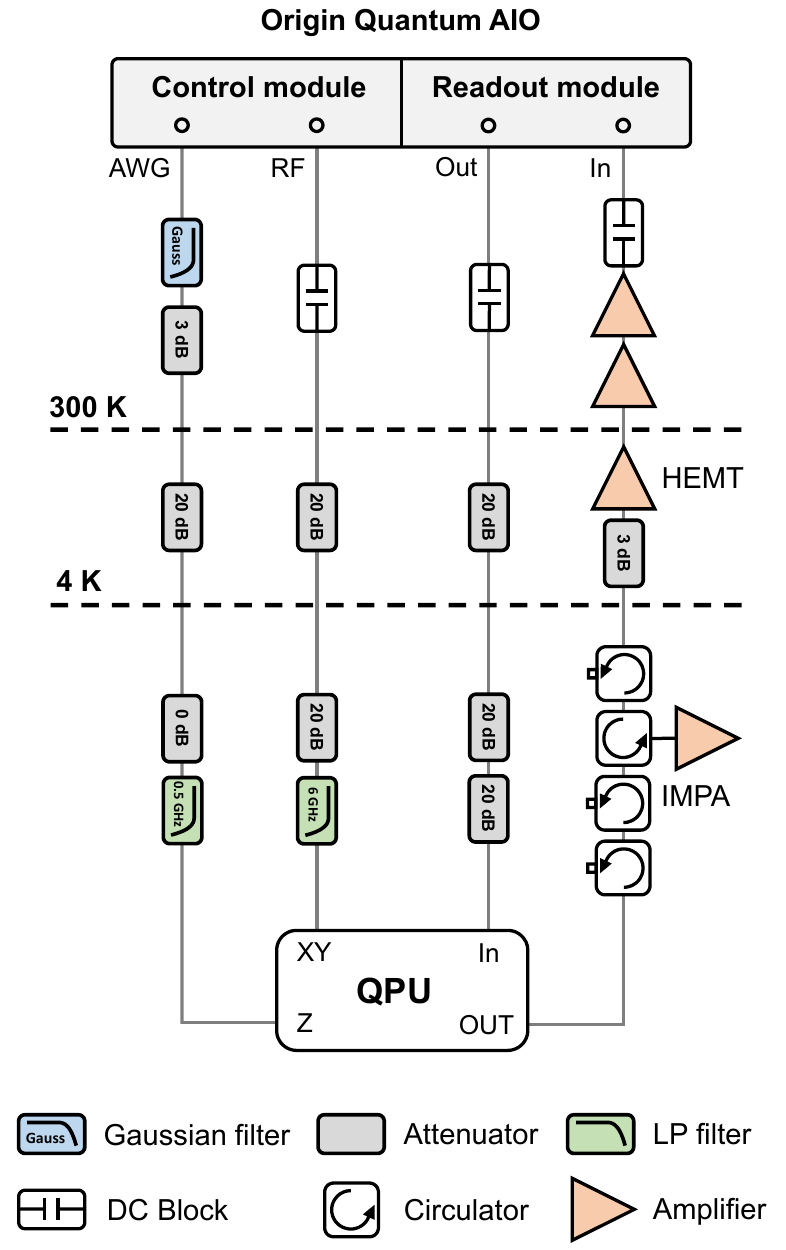}
	\caption{\textbf{Simplified schematic of the quantum processor control and readout wiring.}}
	\label{fig:wiring_diagram}	
\end{figure*}

Within the DR, attenuators and filters are strategically installed at different stages to minimize noise. To suppress thermal noise from higher temperature stages, the XY control lines, Z control lines, and readout input lines incorporate a total attenuation of $40$ dB, $23$ dB, and $60$ dB, respectively. Additionally, all XY control lines are equipped with $6$ GHz low-pass filters, while Z control lines utilize $500$ MHz low-pass filters. DC-blocks are also installed on all XY and readout input lines. To minimize noise introduced through the IMPA control lines, each IMPA pump line is equipped with a $20$ dB attenuator and $12.2\sim 14$ GHz band-pass filter. Each IMPA bias line is equipped with a $20$ dB attenuator and RC filter. The simplified wiring diagram is shown in Fig.~\ref{fig:wiring_diagram}.

\section{Device control calibration}

\subsection{Calibration procedure}

As the number of qubits in a quantum processor increases, calibration becomes increasingly challenging. Traditional one-by-one calibration schemes are time-consuming and lack scalability. More importantly, particularly as parameter drift in both qubits and room-temperature electronics limits the feasibility of prolonged calibration procedures. Furthermore, while the tunability of qubit and coupler frequencies offers great flexibility in designing quantum gates, it simultaneously complicates the optimal selection of working frequencies for single- and two-qubit gates to achieve high parallelism and fidelity.

Here, we briefly introduce our calibration strategy. We define key operational frequencies: the maximum transition frequency of a qubit as the ``sweet point'', the single-qubit gate operating frequency as the ``idle point'', and the CZ gate operating frequency as the ``interaction point''. The calibration process for our processor is structured into five stages: pre-calibration, spectroscopy testing, frequency allocation, fine calibration, and CZ gate calibration.

Even before obtaining complete information about the processor, and particularly when the idle frequencies of the qubits are yet to be determined, we still aim to characterize the processor in a highly parallel manner while simultaneously minimizing the impact of crosstalk and residual coupling. Therefore, during spectroscopy testing, we group the qubits according to the following frequency constraints: (1) The frequency difference between nearest-neighbor qubits, $\Delta f_{\rm{NN}}$, satisfies $|\Delta f_{\rm{NN}}| \geq 300 \ \rm{MHz}$. (2) The frequency difference between next-nearest-neighboring qubits, $\Delta f_{\rm{NNN}}$, satisfies either $\alpha - 60 \ \rm{MHz} \geq |\Delta f_{\rm{NNN}}| \geq 60 \ \rm{MHz}$ or $|\Delta f_{\rm{NNN}}| \geq \alpha + 60 \  \rm{MHz}$, where $\alpha$ is the qubit anharmonicity. Qubits within the same group can be calibrated in parallel, enhancing efficiency. The specific experiments for each calibration stage are detailed below:

\begin{itemize}
\item Pre-calibration. This stage establishes the fundamental parameters of the IMPA and readout resonators: (1) Finding the working point for optimal gain and bandwidth. (2) Determining the frequency of each qubit’s readout resonator. (3) Measuring the readout resonator frequency as a function of the qubit's flux bias. (4) Measuring the readout resonator frequency of the nearest neighbor qubit as a function of the coupler's flux bias. 

\item Spectroscopy testing. Spectroscopy testing begins by biasing all couplers to their maximum frequency as determined in pre-calibration. Simultaneously, a coarse, grouped-parallel calibration of single-qubit gates is performed at the sweet point. This includes: (1) Identifying each qubit's transition frequency. (2) Calibrating the pulse amplitude for $X_{\pi}$ and $X_{\pi/2}$ gates. (3) Measuring and aligning the relative timing between the XY and Z control lines. (4) Evaluating microwave crosstalk between different qubit control lines. Subsequently, we sample $T_1$, $T_2$, and randomized benchmarking (RB) fidelity across each qubit's frequency range. These comprehensive measurements serve as crucial inputs for subsequent frequency allocation algorithms.

\item Frequency allocation. For idle frequency allocation, we use a uniform error model to describe the overall error rate $e_{\rm{total}}$ during parallel execution of single-qubit gates. This model is weighted by factors including $T_1$, $T_2$, XY crosstalk, and residual ZZ coupling, and is expressed as $e_{\rm{total}} = \sum_{i} w_i e_i(f_{\rm{idle}})$, where $e_i$ represents the cost associated with the i-th error mechanism. The weights $w_i$ are determined using a data-driven approach. Specifically, we collect a dataset of measured gate error rates under various frequency configurations and apply open-loop optimization to find the optimal coefficients $w_i$ that best fit the experimental data. For interaction frequency allocation, the error model further considers additional factors such as Z signal distortion and spectator qubit effects. The frequency allocation process employs an open-loop optimization algorithm, which begins with a central qubit and iteratively optimizes the frequency pattern. During each iteration, the total error rate $e_{\rm{total}}^{(t)}$ for gates executed in parallel within a $2$-distance neighborhood is calculated, and the optimal local frequency allocation result from the $t$-th iteration constrains the $t+1$-th iteration, aiming to minimize the overall error rate. The final idle frequency and interaction frequency for the $20$ qubits are shown in Fig.~\ref{fig:1Q_parameters}\textbf{b} and Fig.~\ref{fig:2Q_parameters}.

\item Fine calibration. Following frequency allocation, all couplers are biased to the frequency minimizing ZZ coupling, and all qubits are set to their idle frequencies for parallel fine calibration of single-qubit gates. This stage includes DRAG pulse parameter optimization and readout fidelity optimization (including readout pulse frequency and amplitude, and optimal qubit readout frequency). The single-qubit gate performance measured by simultaneous cross-entropy benchmarking (XEB), shown in Fig.~\ref{fig:1Q_parameters}\textbf{h}.

\item CZ gate calibration. This stage involves preparatory steps: (1) Correction of Z-pulse distortion for both qubits and couplers. (2) Measuring and aligning the relative timing between Z control lines for qubit-qubit and qubit-coupler. Subsequently, CZ gate parameters are calibrated to ensure high-fidelity two-qubit operations, with details provided in the CZ gate calibration section.

\end{itemize}

\subsection{CZ gate calibration}

\begin{figure*}[!th]
	\centering
	\includegraphics[width=1\textwidth]{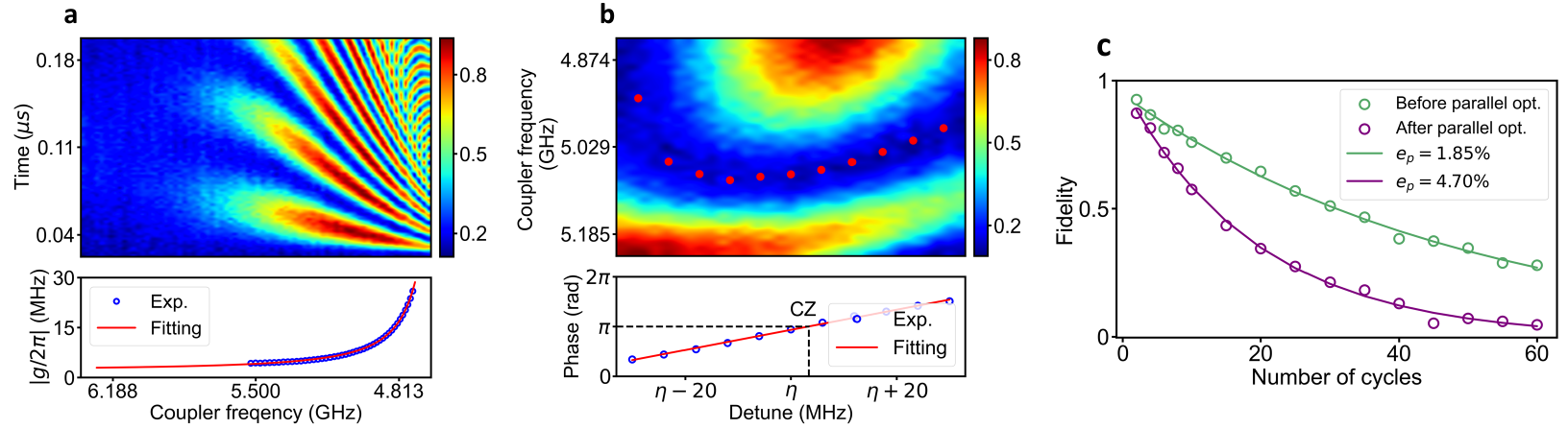}
	\caption{\textbf{Characterization of the parameters of the CZ gate.} $\textbf{a}$, Two-qubit swap experiment shows that the effective coupling strength between neighboring qubits can be dynamically tuned by biasing the intervening coupler. $\textbf{b}$, Top panel: The measured probability of the $|\rm{Q_1} \rm{Q_2} \rangle = |11 \rangle$ state as a function of detuning, while simultaneously varying the coupler frequency, with red dots marking the coupler frequency and detuning at minimum leakage. Bottom panel: The measured controlled phase as a function of detuning, obtained under the parameters corresponding to the red dots in the top panel. The black dashed line indicates the point where the controlled phase is $\pi$. $\eta$ denotes the anharmonicity of $|\rm{Q_1} \rangle$. $\textbf{c}$, Post-optimization via the Nelder-Mead algorithm, the Pauli error rates of parallel two-qubit gates, extracted from simultaneous 2Q-XEB experiments, are significantly reduced.
	}
	\label{fig:2Q_cali}
\end{figure*}

We implement the CZ gate as a coupling-tunable diabatic gate.We consider the dressed states of a three-body system $|\rm{Q_1} \rm{Q_c} \rm{Q_2} \rangle $, the effective coupling strength between $\rm{Q_1}$ and $\rm{Q_2}$ arises from both direct coupling and the contribution mediated by a tunable coupler. This can be rapidly tuned by controlling the coupler frequency, the effective coupling strength can be expressed as:

\begin{eqnarray}\label{eq:S28}
	\widetilde{g} = \frac{1}{2} \left( \frac{1}{\Delta_{{1}}} + \frac{1}{\Delta_{{2}}} - \frac{1}{\Sigma_{{1}}} - \frac{1}{\Sigma_{{2}}} \right) g_{{1,c}} g_{{2,c}} +g_{{1,2}},
\end{eqnarray}
where $\Delta_{i} = \omega_i - \omega_c$, $\Sigma_{i} = \omega_i + \omega_c$ and $g_{i,j}$ is the direct coupling between qubits or couplers. In the Fig.~\ref{fig:2Q_cali}\textbf{a}, we show the measured SWAP dynamics involving the $|11 \rangle \leftrightarrows |20 \rangle $ resonance versus the coupler frequency, which demonstrates that the effective coupling strength can be continuously tuned within the range of $-30$ MHz.

To execute the CZ gate, we apply a fast Z-pulse to the coupler’s control line, tuning it from the coupling-off point to an effective coupling strength of $|g/2\pi| \approx 14 \ \rm{MHz}$. Simultaneously, we bias $\rm{Q_1}$ and $\rm{Q_2}$ from their idle frequency to the interaction frequencies, where $|101 \rangle $ and $|200 \rangle $ are in near resonace with $\rm{Q_2}$ at a lower frequency. After a period of non-adiabatic interaction, this yields a two-qubit gate equivalent to a CZ gate up to a single-qubit phase factor. To minimize rising-edge distortion at the nanosecond scale, both qubit and coupler flux pulses are shaped using smoothed square pulses (flattop pulses with Gaussian edges) defined by error functions:
\begin{eqnarray}\label{eq:S29}
	z(t)=\frac{A}{2}\left(\operatorname{erf}\left( \frac{t-\tau_b}{\sqrt{2}\sigma} \right)-
	\operatorname{erf}\left( \frac{t-\tau_b-\tau_c}{\sqrt{2}\sigma} \right) \right),
\end{eqnarray}
where $A$ is the amplitude of the Z-pulse, $\sigma$ controls the Gaussian smoothing of the rising edge, $\tau_c$ is the effective pulse width, and $\tau_b$ is the buffer at both ends of the effective pulse. The total width of the CZ pulse is $t_{\rm{total}} = \tau_c+2\tau_b$. In the experiment, we set $t_{\rm{total}} = 40$ ns, $\sigma = 1.25 $ and $\tau_b = 7.5$, leaving the parameters of the CZ pulse amplitude $A$ for qubits and coupler to be determined through calibration. Our calibration procedure for the CZ gate is as follows:

\begin{itemize}
	\item First, we sweep the Z-pulse amplitude of the coupler while simultaneously varying the detuning $\Delta$ between $|101 \rangle $ and $|200 \rangle $. This process aims to identify the trajectory that minimizes state leakage, which corresponds to the coupler frequency that maximizes the $|1 \rangle $ population of $\rm{Q_2}$. In Fig.~\ref{fig:2Q_cali}\textbf{b}, red circles represent the coupler frequency fitted from experimental results as a function of detuning ($f_{\rm{coupler}}(\Delta)$) at the minimum leakage. The Z-pulse amplitudes for both the qubits and the coupler in the Control-Phase (CPHASE) gate are then determined using this empirically derived $f_{\rm{coupler}}(\Delta)$ relationship.
	
	\item Next, we measure the controlled phase at various detunings by preparing $\rm{Q_1}$ in $|0 \rangle $ and $|1 \rangle $ respectively, and then performing QST after the CPHASE pulse. As shown by the blue circles in Fig.~\ref{fig:2Q_cali}\textbf{b}, the controlled phase varies continuously within $[0,2\pi]$ as a function of detuning. We perform a linear fit to the data points around $\pi$ to precisely extract the pulse amplitude that yields the desired CZ phase. Subsequently, a virtual Z gate is applied to compensate for any unwanted single-qubit phase accumulated during the CZ pulse.
	
	\item We categorize all CZ gates into two patterns, allowing for parallel execution of CZ gates within each pattern. To optimize the parameters of these parallel CZ gates, we execute a $2$Q-XEB experiment with a fixed depth, and the population of the final state as the objective function. This optimization primarily aims to reduce the impact of residual crosstalk. The optimized parameters include the interaction frequency, the frequency of the coupler, and compensation for single-qubit phases. The Nelder-Mead (NM) algorithm is employed for optimization. As shown in Fig.~\ref{fig:2Q_cali}\textbf{c}, this significantly reduces the simultaneous two-qubit Pauli error. All 19 pairs of optimized two-qubit gate Pauli errors, as measured by simultaneous XEB, are shown in Fig.~\ref{fig:2Q_parameters}\textbf{b} and \textbf{d}.
	
\end{itemize}

\subsection{Readout error mitigation}\label{sec:readout_em}

To correct measurement errors in our superconducting QNN experiments, we adopt the iterative Bayesian unfolding (IBU) method~\cite{nachman2020unfolding}, which addresses the limitations of traditional matrix inversion techniques, such as negative probability entries and amplified statistical uncertainties. We label the measurement response by a $2^n \times 2^n $ assignment (response) matrix $R$, which relates a ideal, error-free probability vector $\boldsymbol{v}$ over computational-basis states to a observed, noisy probability vector $\boldsymbol{w}$ :
\begin{eqnarray}
	w_j= \sum_{i}^{2^{n-1}} R_{ij} v_i,
\end{eqnarray}
where $R_{ij} = \operatorname{Pr}(\operatorname{meas.}=j|\operatorname{prep.}=i)$ is the conditional probability of observing outcome $j$ when the true prepared state was $i$. Under the assumption of negligible readout crosstalk, $R$ can be estimated via single-shot calibration experiments that prepare each computational-basis state and record the measurement distribution. In our setup, we perform these calibration shots once for the full $20$-qubit subset used in classification, and the single-shot readout fidelities are summarized in Fig.~\ref{fig:1Q_parameters}\textbf{f,g}. Given a measured noisy distribution $\boldsymbol{w}$, the IBU update rule proceeds iteratively to estimate the unfolded corrected distribution $\boldsymbol{v}$. Denoting the iteration index by $t$, the update reads:
\begin{eqnarray}
	v^{(t+1)}_i = v^{(t)}_i \sum_{j}^{2^n} \frac{w_j R_{ij}}{\sum_{k} R_{kj} v^{(t)}_k}.
\end{eqnarray}
We initialize $\boldsymbol{v}^0$ to a uniform distribution over states. In practice, the algorithm converges robustly after typically a few tens of iterations. In our experiments, we limit the maximum to $50$ iterations to avoid overfitting or amplifying statistical noise.

In our experiments, the IBU-corrected probabilities are used in subsequent adversarial robustness analysis. We found that IBU effectively suppresses readout biases and yields physically meaningful probability vectors that significantly improve downstream classification calibration and robustness estimation.

\end{document}